\documentclass[twocolumn]{aastex631}
\usepackage{hyperref}
\usepackage{amsmath}
\usepackage{multirow}
\usepackage{ulem}
\usepackage{soul}

\newcommand{\logMs}[1]{{$\log M_{\rm s}/M_\odot#1$}}	
\newcommand{\loglx}[1]{{$\log L_{\rm X}\rm\ [erg\ s^{-1}]#1$}}	
\newcommand{\um}{{$\rm \mu m\ $}}	

\begin{document}

\title{The Stellar Morphology \& Size of X-ray-selected Active Galactic Nuclei Host Galaxies Revealed by JWST}

\author[0000-0003-2213-7983]{Bovornpratch Vijarnwannaluk}
\affiliation{Academia Sinica Institute of Astronomy and Astrophysics (ASIAA), 
11F of Astronomy-Mathematics Building, AS/NTU, No. 1, Section 4, 12 Roosevelt Road, Taipei
106319, Taiwan}
\affiliation{Astronomical Institute, Tohoku University, Aramaki, Aoba-ku, Sendai, Miyagi 980-8578, Japan}

\author[0000-0003-1262-7719]{Zhen-Kai Gao}
\affiliation{Academia Sinica Institute of Astronomy and Astrophysics (ASIAA), 
11F of Astronomy-Mathematics Building, AS/NTU, No. 1, Section 4, 12 Roosevelt Road, Taipei
106319, Taiwan}

\author[0000-0003-2588-1265]{Wei-Hao Wang}
\affiliation{Academia Sinica Institute of Astronomy and Astrophysics (ASIAA), 
11F of Astronomy-Mathematics Building, AS/NTU, No. 1, Section 4, 12 Roosevelt Road, Taipei
106319, Taiwan}

\author[0000-0002-3805-0789]{Chian-Chou Chen}
\affiliation{Academia Sinica Institute of Astronomy and Astrophysics (ASIAA), 
11F of Astronomy-Mathematics Building, AS/NTU, No. 1, Section 4, 12 Roosevelt Road, Taipei
106319, Taiwan}

\author[0000-0001-7713-0434]{Abdurrahman Naufal}
\affiliation{Department of Astronomical Science, The Graduate University for Advanced Studies, 2-21-1 Osawa, Mitaka, Tokyo 181-8588, Japan}
\affiliation{National Astronomical Observatory of Japan, 2-21-1 Osawa, Mitaka, Tokyo 181-8588, Japan}

\author[0000-0001-9882-1576]{Adarsh Ranjan}
\affiliation{Space Telescope Science Institute, 3700 San Martin Drive, Baltimore, MD 21218, USA}

\author[0000-0001-5615-4904]{Bau-Ching Hsieh}
\affiliation{Academia Sinica Institute of Astronomy and Astrophysics (ASIAA), 
11F of Astronomy-Mathematics Building, AS/NTU, No. 1, Section 4, 12 Roosevelt Road, Taipei
106319, Taiwan}

\author[0000-0003-4531-0945]{Chayan Mondal}
\affiliation{Academia Sinica Institute of Astronomy and Astrophysics (ASIAA), 
11F of Astronomy-Mathematics Building, AS/NTU, No. 1, Section 4, 12 Roosevelt Road, Taipei
106319, Taiwan}

\author[0009-0002-9147-9274]{Chih-Yuan Chang}
\affiliation{Academia Sinica Institute of Astronomy and Astrophysics (ASIAA), 
11F of Astronomy-Mathematics Building, AS/NTU, No. 1, Section 4, 12 Roosevelt Road, Taipei
106319, Taiwan}

\author[0000-0002-4205-9567]{Hiddo S.B. Algera}
\affiliation{Academia Sinica Institute of Astronomy and Astrophysics (ASIAA), 
11F of Astronomy-Mathematics Building, AS/NTU, No. 1, Section 4, 12 Roosevelt Road, Taipei
106319, Taiwan}

\author[0000-0002-7026-6782]{Li-Wen Liao}
\affiliation{Institut d’Estudis Espacials de Catalunya (IEEC), Edifici RDIT, Campus UPC, 08860 Castelldefels (Barcelona), Spain}
\affiliation{Institute of Space Sciences (ICE, CSIC), Campus UAB, Carrer de Can Magrans, s/n, E-08193 Barcelona, Spain}

\author[0000-0002-2651-1701]{Masayuki Akiyama}
\affiliation{Astronomical Institute, Tohoku University, Aramaki, Aoba-ku, Sendai, Miyagi 980-8578, Japan}

\author[0000-0001-9970-8145]{Seong Jin Kim}
\affiliation{Institute of Astronomy, National Tsing Hua University, No. 101, Section 2, Kuang-Fu Road, Hsinchu City 30013, Taiwan}

\author[0000-0002-6068-8949]{Shoichiro Mizukoshi}
\affiliation{Academia Sinica Institute of Astronomy and Astrophysics (ASIAA), 11F of Astronomy-Mathematics Building, AS/NTU, No. 1, Section 4, 12 Roosevelt Road, Taipei 106319, Taiwan}

\author[0000-0002-6821-8669]{Tomotsugo Goto}
\affiliation{Institute of Astronomy, National Tsing Hua University, No. 101, Section 2, Kuang-Fu Road, Hsinchu City 30013, Taiwan}

\author[0000-0002-6720-8047]{Yu-Yen Chang}
\affiliation{Department of Physics, National Chung Hsing University, 40227, Taichung, Taiwan}

\author[0000-0002-0930-6466]{Caitlin Casey}
\affiliation{Department of Astronomy and Astrophysics, University of California, Santa Cruz, 1156 High Street, Santa Cruz, CA 95064, USA}
\affiliation{Department of Astronomy, The University of Texas at Austin, 2515 Speedway Blvd Stop C1400, Austin, TX 78712, USA}
\affiliation{Cosmic Dawn Center (DAWN), Denmark}

\author[0000-0001-9187-3605]{Jeyhan S. Kartaltepe}
\affiliation{Laboratory for Multiwavelength Astrophysics, School of Physics and Astronomy, Rochester Institute of Technology, 84 Lomb Memorial Drive, Rochester, NY, 14623, USA}


\author[0000-0003-3596-8794]{Hollis B. Akins}
\affiliation{Department of Astronomy, The University of Texas at Austin, 2515 Speedway Blvd Stop C1400, Austin, TX 78712, USA}

\author[0000-0002-7087-0701]{Marko Shuntov}
\affiliation{Cosmic Dawn Center (DAWN), Denmark}
\affiliation{Niels Bohr Institute, University of Copenhagen, Jagtvej 128, 2200 Copenhagen, Denmark}
\affiliation{University of Geneva, 24 rue du Général-Dufour, 1211 Genève 4, Switzerland}

\author[0000-0002-3560-8599]{Maximilien Franco}
\affiliation{Université Paris-Saclay, Université Paris Cité, CEA, CNRS, AIM, 91191 Gif-sur-Yvette, France}
\affiliation{Department of Astronomy, The University of Texas at Austin, 2515 Speedway Blvd Stop C1400, Austin, TX 78712, USA}

\author[0000-0003-0129-2079]{Santosh Harish}
\affiliation{Laboratory for Multiwavelength Astrophysics, School of Physics and Astronomy, Rochester Institute of Technology, 84 Lomb Memorial Drive, Rochester, NY, 14623, USA}



\begin{abstract}

We investigate the stellar shape and size-mass relationship of X-ray selected Active Galactic Nuclei (AGN) host galaxies using the high-angular resolution and deep sensitivity in the near-infrared of the COSMOS-Web JWST survey field. We present the rest-frame 1-\um size, stellar mass, Sersic index, axis-ratio, Gini-$M_{20}$ parameters of 690 moderate luminosity AGNs between redshift 0-3 and with stellar mass \logMs{\sim 10.75}. We find that AGN host galaxies have effective radius of 1-5 kpc which is in between star-forming (SFG) and quiescent galaxies (QGs) of the same stellar mass. AGN hosts have similar size-mass trends as SFG and QGs, being smaller at higher redshift for the same stellar mass. The slope of the size-mass relationship of AGN host galaxies is steeper than that of star-forming galaxies. Their restframe 1\um  stellar morphology indicates a significant spheroidal component. We observed a low merger fraction ($6\%$) in our sample as well as substructures similar to disks, bars, and spiral arms in the residual images, which are in tension with evolutionary pathways that require major mergers. However, may also be due to the different timescales between mergers and AGN activity.

\end{abstract}



\section{Introduction} \label{sec:intro}

Observations of local galaxies show a broad diversity of shapes, sizes, and colors, which can be broadly grouped into star-forming disk-like galaxies, passive spheroidal galaxies, and irregular galaxies. In the local universe, the morphology of local galaxies depends strongly on stellar mass, with spheroidal galaxies being more massive than disky galaxies  \citep{2006MNRAS.373.1389C,2009MNRAS.393.1324B,2016MNRAS.457.1308M}. Beyond the local universe, observations of the size-mass relationship of galaxies show two major trends. First, the size of galaxies increases with the stellar mass. Second, at a fixed stellar mass ($M_{s}$), the size of galaxies decreases towards the early universe. Moreover, observations also show a difference between the size of star-forming galaxies (SFG) and quiescent galaxies (QG). They show that among massive galaxies with stellar mass above \logMs{\geq 10}, quiescent galaxies are more compact than star-forming galaxies at the same stellar mass \citep{2012MNRAS.421.1007K,2014ApJ...788...28V,2024ApJ...972..134M,2024ApJ...964..192I,2024ApJ...962..176W,2024ApJ...960...53V,2025arXiv250407185Y}. 
 
Several pathways of morphological transformation have been proposed to explain the evolution and the build-up of the stellar mass of these massive galaxies. Overall, a dissipative process which results in a loss of angular momentum is required to triggers compaction of gas \citep{2009ApJ...703..785D,2014MNRAS.438.1870D,2015MNRAS.450.2327Z,2015MNRAS.447.3291C,2002ApJS..140..303L}. Gas-rich major mergers provide one pathway to produce these massive spheroidal galaxies \citep{1991ApJ...370L..65B,2005ApJ...620L..79S,2006ApJS..163....1H,2008ApJS..175..390H,2016MNRAS.463.3948D,2023MNRAS.522.4515L}. Simulations show that following a merger, large inflows of gas are funneled in to the center of galaxies, which increases star formation. The violent gravitational forces destroy the disks, and what remains is a spheroidal galaxy. Besides mergers, secular processes such as violent disk instabilities, cold streams, or counter-rotating disks can also result in compaction in the high-redshift gas-rich universe\citep{1999ApJ...514...77N,2008ApJ...688...67E,2016MNRAS.456.2052I,2015MNRAS.450.2327Z,2015MNRAS.449.2087D}.

Within the morphological transition of these models, an active galactic nuclei (AGN) is expected to be triggered due to the inflow of gas \citep{2006ApJS..163....1H,2006ApJS..166....1H,2011ApJ...741L..33B,2018MNRAS.478.3056B}.  Hydrodynamic simulations show that AGN feedback can prevent the resettling of gas into a new star-forming disk \citep{2016MNRAS.463.3948D}. This prevents the formation of a new rotation-dominated star-forming disk in the post-compaction phase, allowing the galaxy to maintain the dispersion-dominated shape. AGN activity is also one of the quenching mechanisms to stop the compact star formation, hindering in-situ stellar mass growth. Hence, the additional stellar mass must be obtained ex-situ via minor mergers.

Investigating the size, mass, and shape of AGN host galaxies in comparison with star-forming (SFG) and quiescent galaxies (QG) could provide insight into how AGN activity and triggering fit into the process of morphological transformation and size evolution of galaxies. However, such analysis has historically been difficult due to the requirement of multiple high angular resolution observations and the low number density of AGN, which affects sample statistics. Recent efforts with large area optical imaging surveys and Hubble Space Telescope (HST) imaging show that AGN host galaxies have smaller sizes than star-forming galaxies, but not as small as QGs \citep{2019ApJ...887L...5S,2021ApJ...918...22L,2024MNRAS.527.4690L}. Previous HST studies also show that most AGN reside in disk galaxies and not disturbed systems \citep{2011ApJ...727L..31S,2011ApJ...726...57C,2012ApJ...744..148K,2019ApJ...877...52Z}. With the recent imaging from the James Webb Space Telescope (JWST), studies are shedding more light on the relationship between AGN and host galaxies' properties, with some focusing on the relationship between stellar mass and the supermassive blackhole masses \citep{2024ApJ...962..139Z,2025ApJ...979..215T,2025arXiv250503876D}, as well as on the shapes and sizes of AGN host \citep{2022ApJ...939L..28D,2024ApJ...962...93Z,2025ApJ...978...74B,2025arXiv251011010B,2025arXiv250503876D}.

In this study, we used data from the COSMOS-Web survey \citep{2023ApJ...954...31C} combined with the Chandra COSMOS Legacy Survey (CCLS; \citealt{2016ApJ...819...62C,2016ApJ...817...34M}) to investigate the size and shapes of AGN host galaxies.  Among all the current JWST imaging surveys, this combination provides the largest sample of AGN host galaxies with high angular resolution near-infrared observations in multiple bands.  In section~\ref{sec:data}, we describe our reduction of the JWST datasets. In section~\ref{sec:anal}, we described our analysis methodology, including AGN host decomposition, single Sersic fitting, and non-parameteric morphologies. In section~\ref{sec:results}, we describe our results on the shape and size of AGN host galaxies compared to NUVrJ selected starforming galaxies and quiescent galaxies (QGs). Here, we adopt flat standard $\rm \Lambda CDM$ cosmology with $H_{\rm 0} = 70\ \rm km\ s^{-1}\ Mpc^{-1}$,$\Omega_{\rm M} = 0.3$, and $\Omega_{\rm \lambda} = 0.7$ was assumed. Magnitudes are reported in the AB magnitude system \citep{1983ApJ...266..713O}.

\section{Data} \label{sec:data}
\subsection{NIRCam Imaging Data} \label{sec:data:imaging}
The COSMOS-Web survey is a JWST treasury program that surveys the central $\sim 0.6\ \rm Deg^2$ region of the multiwavelength COSMOS Survey \citep{2007ApJS..172....1S}. The survey was performed using 4 NIRCam and 1 MIRI filters, including F115W, F150W, F277W, F444W, and F770W.  It also overlaps with the smaller Public Release IMaging for Extragalactic Research survey (PRIMER, \citet{2021jwst.prop.1837D}), which is centered on the CANDELS-COSMOS field and has a deeper imaging depth and an additional 4 NIRCam and 1 MIRI filter, including F090W, F200W, F356W, F410M, and F1800W. 

In this work, we used the NIRCam imaging datasets, which were reduced using our data reduction pipeline (version 13). The pipeline is built on top of the official NIRCam imaging pipeline (version 1.13.4) and uses PMAP context version 1237. Our pipeline follows the same structure with three major stages, but with additional processing steps to deal with noise, bad pixels, and image artifacts, which differ from the official pipeline and recently released COSMOS 2025 datasets \citep{2025arXiv250603256F,2025arXiv250603243S,2025arXiv250603306H}.

\subsubsection{Pipeline Stages}

In Stage 1, we first apply the standard imaging pipeline ($\rm \texttt{calwebb\_detector1}$) up to the ramp fitting step to reduce the NIRCAM images. It is known that NIRCAM images suffer from cosmic ray hits (snowballs) and persistence from bright stars in previous exposures. To mitigate these problems, we apply additional artifact removal steps in each pipeline stage. Before entering the ramp fitting step, we use the snowball masking algorithm ({$\texttt{snowblind}$})\footnote{\url{https://github.com/mpi-astronomy/snowblind}} to mask the region affected by snowballs.  We construct a lookup table that records the time and the positions of the saturated pixels to record pixels affected by persistence. For each uncal image, we reject groups that have more than 100,000 pixels flagged as jump pixels. Afterward, we run the up-the-ramp fit to produce the rate files. 

In stage 2, we apply the standard stage 2 pipeline ($\texttt{calwebb\_image2}$) but skip the distortion correction steps to minimize the number of resampling performed on the data. The distortion correction will be applied at the end as part of the resampling step to produce the final mosaic. The final product of this stage is the distortion-uncorrected flux-calibrated images, which will be provided to stage 3.

\begin{figure}
    \centering
    \includegraphics[width=\linewidth]{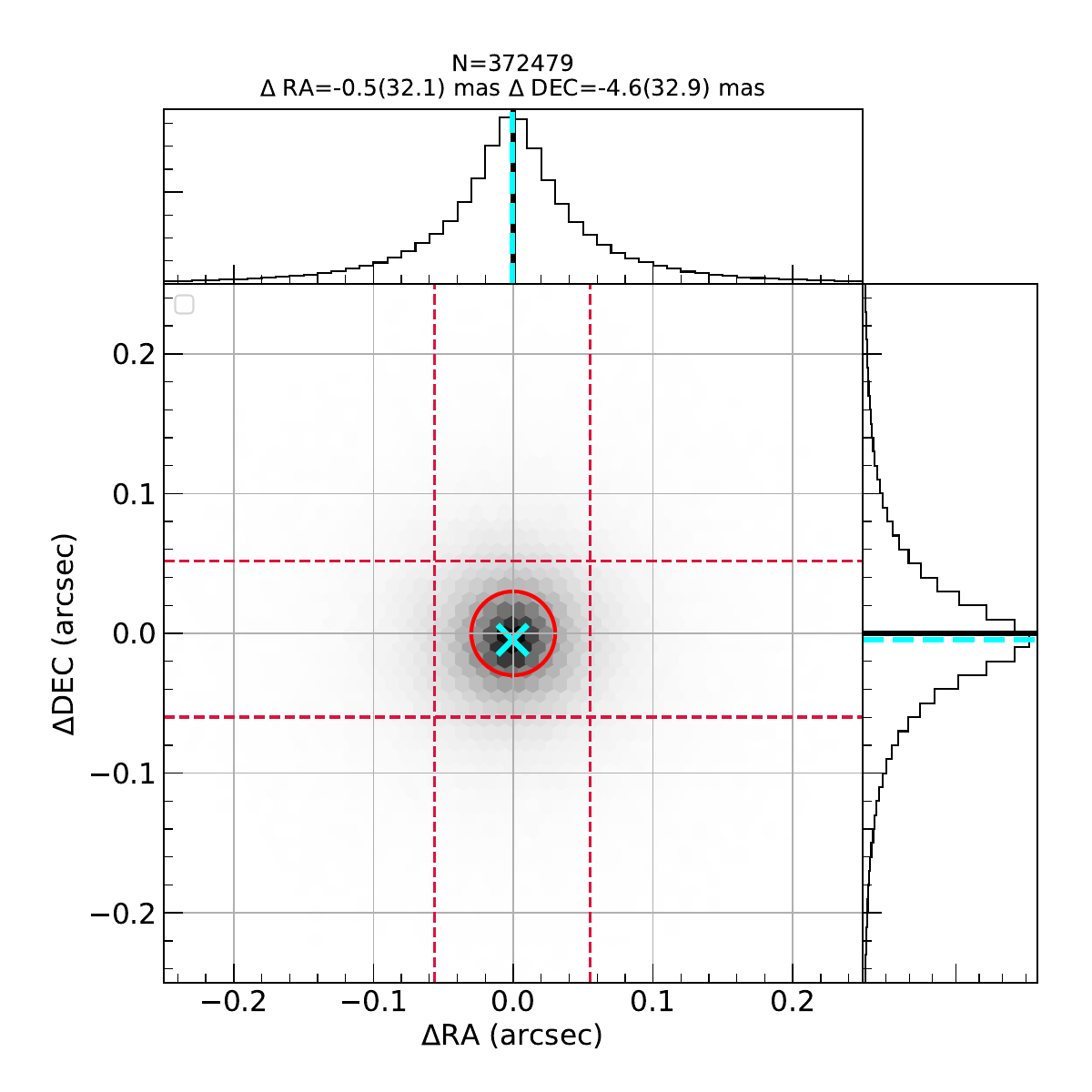}
    \caption{Deviation in the RA and declination of the JWST detected galaxies against the coordinates presented in the HSC-PDR3 catalogs. The red dashed lines show the 1$\sigma$ deviation, while the cyan cross shows the median shift from 0. The red circle shows 1 pixel size (0.03").  The numbers in parentheses show the median absolute deviation (MAD) in RA and declination.}
    \label{fig:astrometry}
\end{figure}

In stage 3, we apply the standard stage 3 pipeline ($\texttt{calwebb\_image3}$). This stage performs the absolute astrometry refinement, additional artifact removal, and image resampling \& image combination. We perform the astrometric alignment using $\texttt{TweakReg}$, which can perform astrometric refinement on distortion-uncorrected images given that the distortion correction table is available in the image header. Due to the lack of faint GAIA stars in the narrow NIRCam field, we align the astrometry to that of the Hyper-Suprime Cam Public Data Release 3 (HSC-PDR3, \citealt{2022PASJ...74..247A}) Deep survey in the COSMOS field. The HSC astrometric system is defined following that of the PANS-STARR1 survey \citep{2016arXiv161205560C}. For each image, we obtain $\sim 100-200$ detections with a WCS fitting RMS of $\sim 0.1$ pixel, which is sufficient to produce mosaics. Internal consistency checks of the astrometry are shown in Figure~\ref{fig:astrometry}. We perform consistency checks against the GAIA DR3 catalog \citep{2023A&A...674A...1G}, which shows a small $\sim 0.5$ pixel shift against the GAIA astrometric system. This shift is likely due to differences between the PAN-STARRS and GAIA astrometric systems, but does not affect the results of this work. The GAIA astrometric system will be adopted in future iterations of the data reduction. Here, we correct these shifts on the catalog level. 

During this stage, the look-up table produced in stage-1 is used to flag bad pixels from being used in subsequent background and pedestal measurements. We also subtract the $1/f$ noise from the images during this stage. Here, we measure the $1/f$ noise after subtracting the local background. This procedure differs from the standard pipeline but reduces the instances of over-subtraction of  $1/f$ noise near bright diffuse galaxies.

\subsubsection{Wisp Removal}

NIRCam suffers from scattered light artifacts, especially in the short bands  (F090W, F115W, F150W, and F200W) known as wisp. We apply an additional step to our pipeline based on the algorithm presented by \citet{2023PASP..135h5003R} to remove them. The algorithm is a two-step process. First, a reference mosaic with no image artifacts is made, generally with the longest wavelength band. Second, pixels with extremely blue colors in the shorter bands due to the presence of scattered light are flagged.  We refer to \citet{2023PASP..135h5003R} for the full description of the artifact removal procedure.  Here, we visually examined each F444W image and flagged regions that have image artifacts by hand to construct a reference mosaic. We apply the algorithm to the shorter bands to flag areas affected by wisp so that they are not used to produce the final mosaic. After these steps, we produce the mosaics with a pixel scale of 0.03 arcsecond per pixel using the artifact-cleaned image. Pixels that were flagged by previous steps were not used in this process

\subsubsection{Point Spread Function (PSF) Model Construction}

We used PSFex \citep{2013ascl.soft01001B} to construct the effective point spread function (PSF) in each band. PSFex is a program that constructs the PSF models' image cutouts from SExtractor \citep{1996A&AS..117..393B}. First, we run SExtractor on each mosaic image to create the initial candidate list. Second, we select the candidates based on their magnitude and half-light radius. For COSMOS-Web, we select PSF candidates that have a magnitude between 20-24 and a half-light radius between 1-2 pixels for F115W and F150W, while for F277W and F444W, we choose candidates with a half-light radius between 2.0-3.0 and 2.5-3.5 pixels, respectively. For PRIMER-COSMOS, we select PSF candidates that have a magnitude between 20-26 and a half-light radius between 1-1.75 pixels for F090W and F200W, while for F356W and F410M, we choose candidates with a half-light radius between 2.0-3.0 pixels. The faint magnitude cutoff was applied to minimize contamination from faint compact galaxies whose half-light radius converges with a faint point source. 

Third, we perform an additional source extraction for each candidate cutout to detect nearby sources and mask them from the cutout. The regions with detection other than our PSF candidate were replaced with Gaussian noise determined from the RMS of the background. The cutout is discarded if the fraction of non-background pixels exceeds 40\% of the cutouts. We fit a 2D Gaussian to estimate the axis ratio and offset from the image center. We limit the candidate sample to those with axis-ratio $>0.9$, which are less than 1 pixel offset from the image center. This step ensures that no sources that are blended with nearby objects are passed to PSFex. In this procedure, a few thousand PSF candidates remain for each in the COSMOS-Web bands, whereas the PRIMER bands have a few hundred PSF candidates.

\begin{figure}
    \centering
    \includegraphics[width=\linewidth]{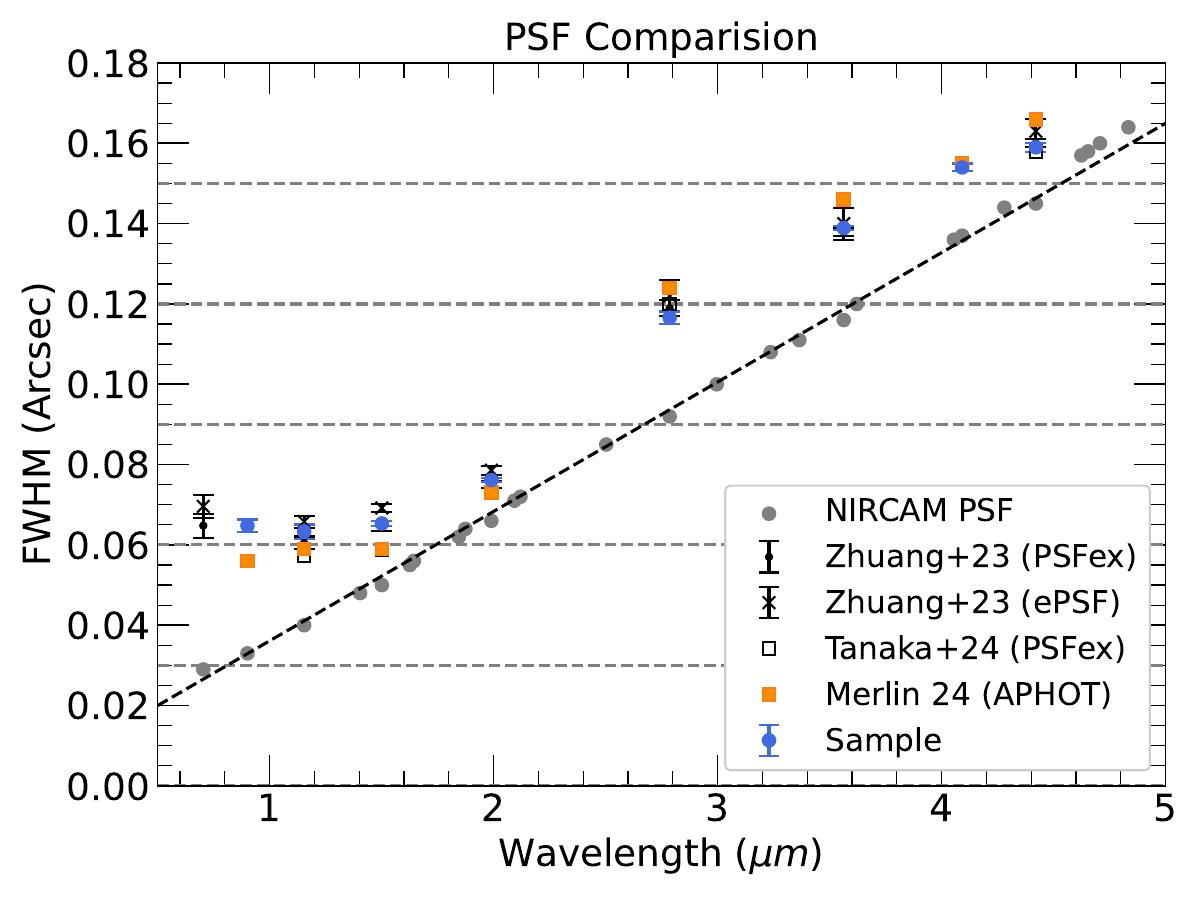}
    \caption{Measured FWHM of each NIRCam band in our mosaic compared with the empirical FWHM and recent analysis of 
 NIRCam data from other studies. The horizontal dashed lines show steps of 1 pixel. The diagonal line shows the diffraction limit FWHM.}
    \label{fig:fwhm}
\end{figure}

In addition, it is known that the variations in the position angle (PA) during the observations create PSFs at different angles in the image. This effect is seen in the 2 epochs of the COSMOS-Web observations. To account for this dependence, we separate the PSF candidate objects based on their epoch of observations, including epoch-1 (south), epoch-2 (north), and PRIMER regions. The PSF candidates catalog is written out as a new FITS LDAC catalog for PSFex. We create PSF models with sizes of 101 and 201 pixels with oversampling of 1 and sizes of 201 and 401 pixels for oversampling of 2, normalized to unity. Table~\ref{tab:psfcand} shows the final full-width half maximum (FWHM), 50\% flux enclosing radius, and 80\% flux enclosing radius for each filter. Figure~\ref{fig:fwhm} shows the effective full-width half maximum (FWHM) of the PSF in each band compared to reported PSF FWHM\footnote{https://jwst-docs.stsci.edu/jwst-near-infrared-camera/NIRCam-performance/NIRCam-point-spread-functions} and NIRCam PSF measurements from other studies. Our result follows a similar trend with other teams, which shows that the effective FWHM can be larger than diffraction-limited, with the effect most apparent in the shortest wavelengths \citep{2024ApJ...962..139Z,2025ApJ...979..215T,2024A&A...691A.240M}.

\begin{deluxetable}{lccc}[!htb]
\tablecaption{PSF Full-width Half Maximum, 50\% Encircled Energy ($f_{50}$) and 80\% Encircled Energy ($f_{80}$) in Each Band }
\label{tab:psfcand}
\tablehead{
    \colhead{Filter} & FWHM & $f_{50}$ & $f_{80}$ \\ 
    \colhead{} & Arcsec & Arcsec & Arcsec
    }
    \startdata
    F090W & $0.065 \pm 0.002$ & $0.054 \pm 0.008$ & $0.215 \pm 0.109$ \\
    F115W & $0.063 \pm 0.002$ & $0.050 \pm 0.004$ & $0.175 \pm 0.020$ \\
    F150W & $0.065 \pm 0.001$ & $0.048 \pm 0.002$ & $0.147 \pm 0.016$ \\
    F200W & $0.076 \pm 0.001$ & $0.053 \pm 0.001$ & $0.143 \pm 0.008$ \\
    F277W & $0.117 \pm 0.002$ & $0.079 \pm 0.002$ & $0.185 \pm 0.007$ \\
    F356W & $0.139 \pm 0.001$ & $0.088 \pm 0.001$ & $0.207 \pm 0.008$ \\
    F410M & $0.154 \pm 0.001$ & $0.096 \pm 0.003$ & $0.234 \pm 0.021$ \\
    F444W & $0.159 \pm 0.001$ & $0.099 \pm 0.001$ & $0.246 \pm 0.006$ \\
    \enddata
\end{deluxetable}

\subsection{JWST Photometric Catalog} \label{sec:data:jwstphot}

We perform PSF-matched aperture photometry on the PSF-matched mosaic images. First, we construct the convolution kernels to convolve the PSF of each band to that of the F444W band using the deconvolution code PyPher \citep{2016A&A...596A..63B}. Then we convolve all using the kernels. Second, we perform aperture photometry using the code Aperpy \citep{2024ApJS..270....7W}. Aperpy is an aperture photometry pipeline that uses Source Extractor in Python \citep{2016JOSS....1...58B}, which was built on the core libraries of SExtractor. Before we construct the detection image and perform the photometry, we perform 2D background subtraction by using 10x10 pixel grids and perform $3\sigma$ clipping to reject outlying boxes.

The detection image used in our photometry is an inverse-variance weighted combination image of the F277W and F444W mosaics. We chose the F277W and F444W images because they are the deepest and least affected by the intrinsic dust attenuation in the observed galaxies. We set the detection limit to 1.2 sigma detection with 5 connecting pixels. For circular aperture photometry, we use 0.32, 0.48, 0.70, 1.00, and 1.4 arcsecond diameter apertures. For Kron photometry, we set the multiplicative factor to the kron radius as 2.5. We also perform adaptive "auto" photometry, which chooses between circular aperture and kron aperture photometry. Circular aperture photometry is chosen when the circularized Kron radius is smaller than the smallest circular aperture.

Image resampling introduces correlated noise, which artificially reduces the noise in the error maps. To correct for this effect, we perform an empty aperture analysis \citep{2012A&A...545A..23B,2016ApJS..224...24L} to correct for the effect of pixel-pixel correlation. The correction factor to the photometry is defined as  $k_{aper}=\sigma_{\rm sky}/\sigma_{\rm phot}$. We first divide the mosaic image into 20x20 grids. For each grid, we place 3000 circular apertures on empty parts of the image and measure the standard deviation of the sky background ($\sigma_{\rm sky}$) and photometric uncertainty ($\sigma_{\rm phot}$). The survey depth was calculated in a grid of 0.5 arcminutes wide boxes. A $3\sigma$ clip was applied to remove apertures that may fall on remaining bad pixels or overlap with faint objects. The correction factors for all apertures are shown in table~\ref{tab:pixpixcorr}.  We adopt the correction factor of the largest aperture to perform the correction to the total photometry. 

\begin{figure}
    \centering
    \includegraphics[width=\linewidth]{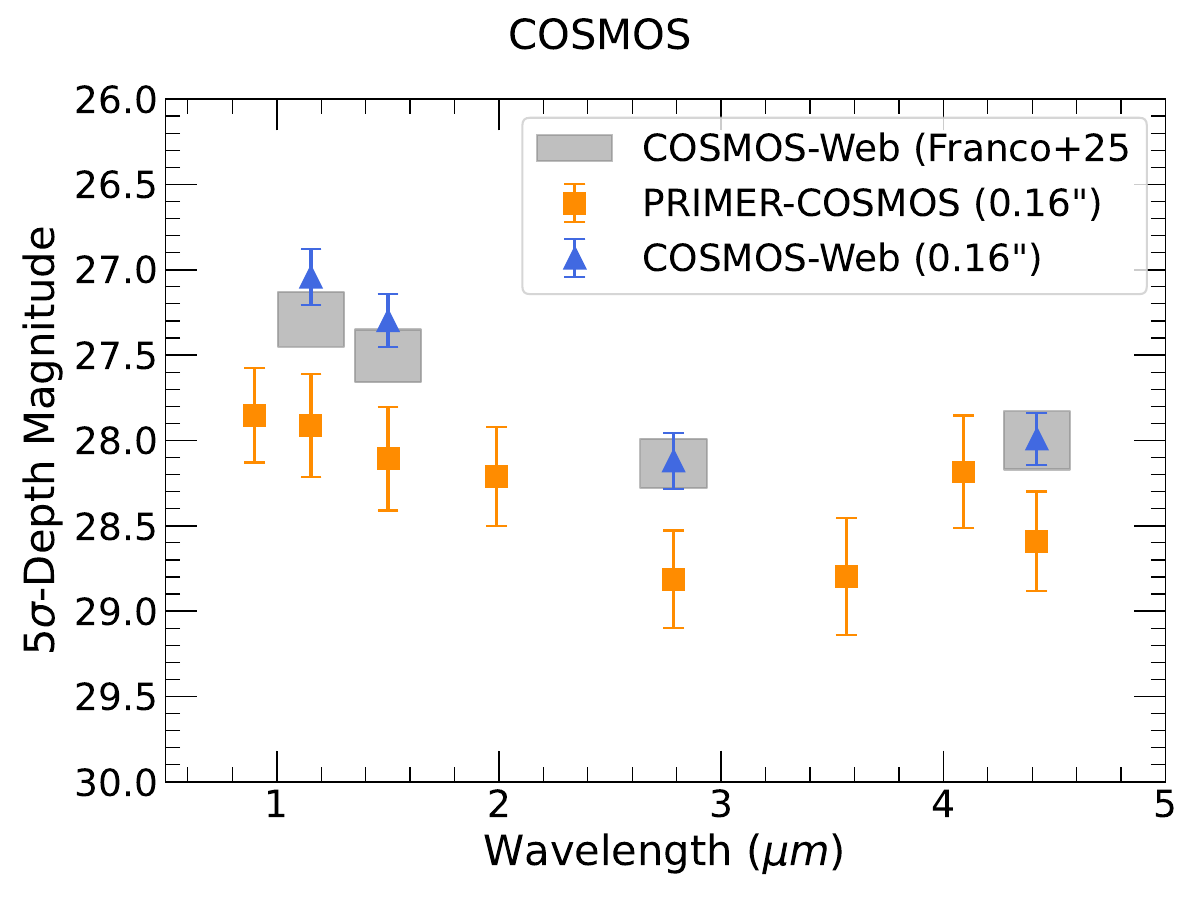}
    \caption{$5\sigma$ depth in a 0.16" diameter blank sky aperture in the COSMOS-Web and PRIMER COSMOS regions shown with blue triangles and orange squares, respectively.  Grey shaded regions show the 5-$\sigma$ depth in a 0.15 arcsecond diameter aperture of the COSMOS 2025 JWST imaging dataset \citep{2025arXiv250603256F}. The upper edge and lower edge of the shaded area represent the depth with 2 and 4 overlapping images. }
    \label{fig:survey_depth}
\end{figure}

The empty apertures were also used to measure the $5\sigma$ depth in the mosaic. Table~\ref{tab:pixpixcorr} presents the 5$\sigma$ depth within a 0.15" diameter aperture of each mosaic. The COSMOS-Web survey area also contains the deeper PRIMER-COSMOS survey, which also contains the additional F090W, F200W, F356W, and F410M bands. Figure~\ref{fig:survey_depth} shows the measured survey depth in the COSMOS-Web area and the PRIMER-COSMOS area compared to the UltraVista photometry in the deep and ultra-deep regions from the COSMOS2020 catalog \citep{2022ApJS..258...11W}. The NIRCam observations are $\sim 1$ magnitude deeper than the COSMOS2020  catalog, with the PRIMER region deeper by $\sim 1.5$ magnitude.

\begin{deluxetable*}{lcccccc}[!htb]
\tablecaption{Pixel-Pixel Correlation Factors \& $5\sigma$ Survey Depth in 1" Diameter Aperture}
\label{tab:pixpixcorr}
\tablewidth{0pt}
\tablehead{
    \colhead{Filter} & \multicolumn5c{Flux Correction Factor} & Survey Depth \\ 
     {}  & 0.32" & 0.48"  & 0.50"  & 1.00" & 1.40" & AB Mag } 
\startdata
& \multicolumn{5}{c}{COSMOS-Web} & \\
F115W & $ 1.34 \pm 0.18 $ & $ 1.52 \pm 0.22 $ & $ 1.74 \pm 0.27 $ & $ 1.98 \pm 0.34 $ & $2.26 \pm 0.42 $ & $25.38 \pm 0.17$ \\
F150W & $ 1.33 \pm 0.18 $ & $ 1.52 \pm 0.23 $ & $ 1.74 \pm 0.30 $ & $ 2.00 \pm 0.37 $ & $2.29 \pm 0.46 $ & $25.62 \pm 0.17$ \\
F277W & $ 2.39 \pm 0.28 $ & $ 2.68 \pm 0.33 $ & $ 3.01 \pm 0.41 $ & $ 3.37 \pm 0.50 $ & $3.74 \pm 0.60 $ & $26.51 \pm 0.17$ \\
F444W & $ 2.13 \pm 0.23 $ & $ 2.33 \pm 0.28 $ & $ 2.55 \pm 0.33 $ & $ 2.79 \pm 0.38 $ & $3.06 \pm 0.46 $ & $26.46 \pm 0.16$ \\
& \multicolumn{5}{c}{PRIMER-COSMOS} &  \\
F090W &$1.38 \pm 0.12$ & $1.57 \pm 0.17$ & $1.82 \pm 0.22$ & $2.12 \pm 0.30$ & $2.44 \pm 0.39 $ & $26.14 \pm 0.26 $ \\
F115W &$1.48 \pm 0.17$ & $1.71 \pm 0.20$ & $2.00 \pm 0.31$ & $2.37 \pm 0.37$ & $2.72 \pm 0.46 $ & $26.17 \pm 0.28 $ \\
F150W &$1.49 \pm 0.19$ & $1.75 \pm 0.23$ & $2.06 \pm 0.31$ & $2.45 \pm 0.41$ & $2.91 \pm 0.55 $ & $26.35 \pm 0.29 $ \\
F200W &$1.45 \pm 0.15$ & $1.69 \pm 0.23$ & $2.01 \pm 0.32$ & $2.38 \pm 0.42$ & $2.82 \pm 0.54 $ & $26.42 \pm 0.27 $ \\
F277W &$2.74 \pm 0.30$ & $3.15 \pm 0.40$ & $3.58 \pm 0.52$ & $4.08 \pm 0.66$ & $4.62 \pm 0.78 $ & $27.15 \pm 0.28 $ \\
F356W &$2.58 \pm 0.24$ & $2.89 \pm 0.31$ & $3.28 \pm 0.42$ & $3.77 \pm 0.53$ & $4.19 \pm 0.67 $ & $27.16 \pm 0.32 $ \\
F410M &$2.44 \pm 0.21$ & $2.70 \pm 0.26$ & $2.98 \pm 0.32$ & $3.30 \pm 0.40$ & $3.67 \pm 0.49 $ & $26.62 \pm 0.32 $ \\
F444W &$2.42 \pm 0.25$ & $2.68 \pm 0.31$ & $3.00 \pm 0.39$ & $3.37 \pm 0.49$ & $3.63 \pm 0.55 $ & $27.02 \pm 0.29 $ \\
\enddata
\end{deluxetable*}

\subsection{Chandra X-ray Catalog} \label{sec:data:ccls}

The C-COSMOS Legacy survey (CCLS; \citealt{2016ApJ...819...62C,2016ApJ...817...34M}) is a deep X-ray survey performed by the \textit{Chandra Space Telescope} using the ACIS-I imager in the 0.5-2 (soft), 2-10 (hard), and 0.5-10 (full) keV energy range in the COSMOS survey. It expands the original 1.7 deg$^2$ C-COSMOS survey \citep{2009ApJS..184..158E} to cover the whole 2.2 deg$^2$ area of the COSMOS field with uniform depth. The survey depths in 0.5-2 keV, 2-10 keV, and 0.5-10 keV bands are $2.2\times10^{-16}$, $1.5\times10^{-15}$, and $8.9\times10^{-16}$ $erg\ cm^{-2} s^{-1}$, respectively. In total, the dataset contains 4016 X-ray sources with 3814, 2920, and 2440 sources detected in the full, soft, and hard bands, respectively.  Optical infrared (OIR) counterparts were presented by \citet{2016ApJ...817...34M} by matching the X-ray source catalogs to the COSMOS2015 \citep{2016ApJS..224...24L}, i-band \citep{2007ApJS..172...99C}, IRAC 3.6 \um \citep{2007ApJS..172...86S} catalogs, respectively. Photometric redshift analysis for all the X-ray sources was estimated using AGN templates following the procedure outlined in \citet{2009ApJ...690.1250S,2011ApJ...742...61S}. We use this X-ray dataset for our initial AGN sample selection. However, we perform a rematching of the OIR catalogs to the latest COSMOS2020  \citep{2022ApJS..258...11W} catalog. This is because the previous catalog was matched with the COSMOS2015 catalog with a pre-GAIA astrometric reference and not all X-ray sources have an OIR counterpart. We describe the process of rematching the X-ray, OIR counterparts, and JWST detection in the next section (~\ref{sec:data:agnsample}).

\subsection{AGN Sample Selection} \label{sec:data:agnsample}

In this section, we describe the selection of AGN used in this work and how the sources were initially matched to the COSMOS2020  catalog, and then to the JWST catalog. First, we select X-ray sources that fall in the footprint of the COSMOS-Web survey using the X-ray source coordinates. Among the 4016 X-ray sources in the CCLS Survey, 1250 sources are in the COSMOS-Web footprint\footnote{The corners are (150.57994, 2.4524263), (149.92938, 2.6918347),(149.66290, 1.9614567), and (150.31471, 1.7269094), respectively}. The large difference is due to wider coverage in CCLS, which covers the whole cosmos field. For each X-ray source, \citet{2016ApJ...817...34M} presented OIR counterparts from COSMOS2015 Ultravista K-band, COSMOS i-band, or IRAC 3.6\um , survey. We choose the latter counterpart only if the prior ones are not available.  

\begin{figure*}
    \centering
    \includegraphics[width=\linewidth]{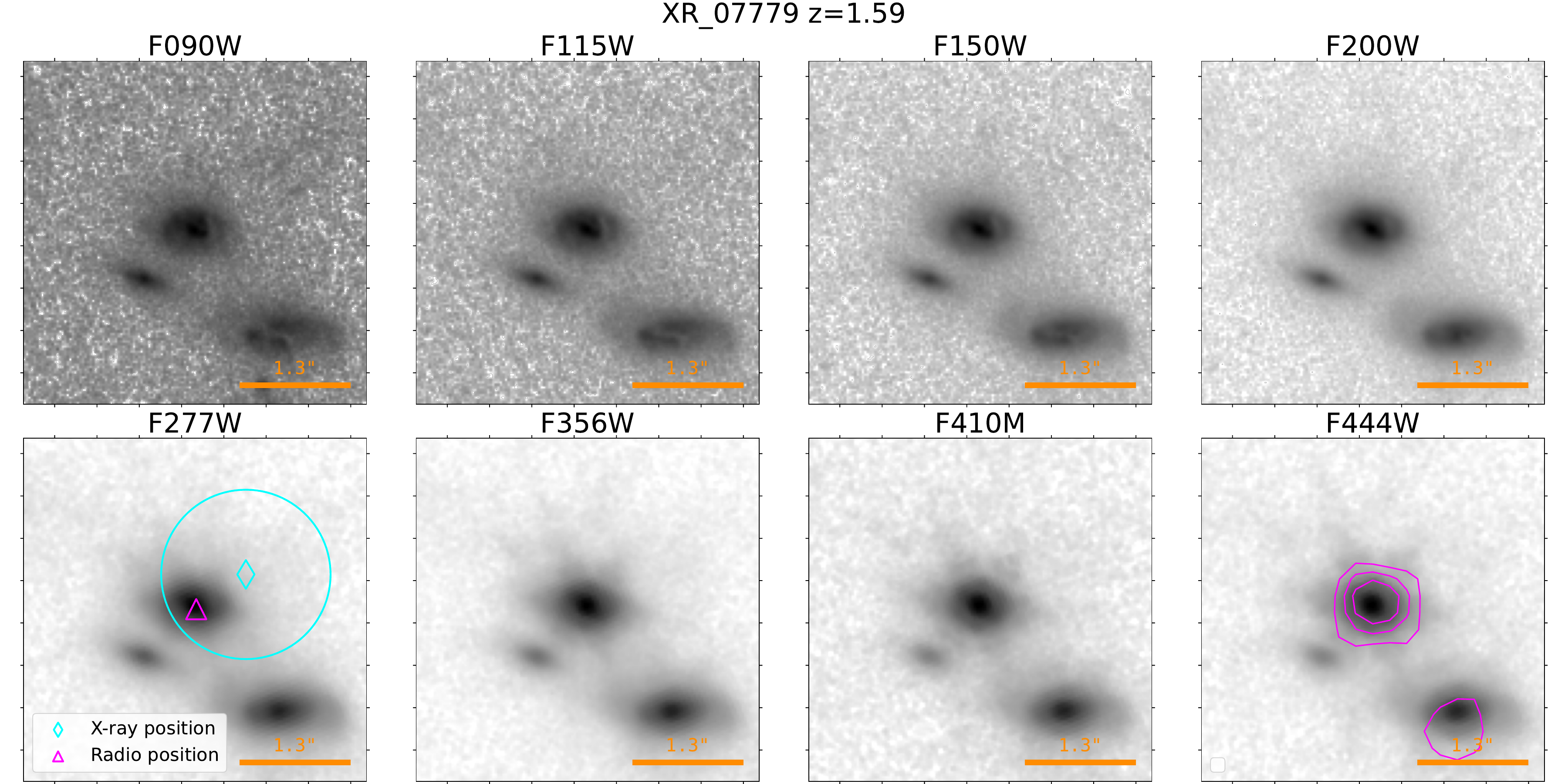}
    \caption{Example Cutout images of one of our targets in each band. The X-ray and radio source position is shown in cyan diamonds and magenta triangles, respectively. The magenta contours in the F444W cutout image show the signal in the radio detection. The cyan circle shows the maximum 1" positional uncertainty of Chandra. The contours are at $3,5 and 7\sigma $ level. }
    \label{fig:enter-label}
\end{figure*}

Second, we match the X-ray source position to the radio VLA 3 GHz positions \citep{2017A&A...602A...1S} using a 1$''$ radius aperture. This cross-matching was performed because visual inspection of the X-ray and radio source position and the JWST image shows that the X-ray position sometimes can show offsets from the JWST counterpart, while the radio source detection shows no random offset at all. The shifts are possibly due to the positional accuracy of Chandra, which has an uncertainty of $\sim 0.6$ arcsecond but can be as high as 1 arcsecond.\footnote{\url{https://cxc.cfa.harvard.edu/proposer/POG/html/chap5.html\#tth\_fIg5.5}} By matching with the radio sources, we were able to leverage the better positional accuracy of radio telescopes, which we confirmed by visual inspection of the radio detection and the JWST images. 

Among the X-ray sources, 388 have radio detection. For these sources, we adopt 366 OIR counterparts from the VLA 3 GHz multiwavelength catalog \citep{2017A&A...602A...2S} and adopt 21 OIR counterparts from the CCLS multiwavelength catalog.  For 862 X-ray sources without radio detection, we first adopt 855 OIR counterparts from the CCLS multiwavelength catalog. For the remaining 8 sources with no OIR counterparts in either the VLA 3 GHz catalog or multiwavelength catalog,  we assume that the position of the OIR counterpart is at the radio source position if there is a radio detection; otherwise, we assume that the counterpart is at the X-ray source position. In summary, among the 1250 X-ray sources, 1242 have secure OIR counterparts, whereas 8 have OIR positions based on either X-ray or radio positions.

Before matching with our JWST photometric catalog, we match our X-ray sources to the COSMOS2020  catalog using a 1-arcsecond radius. Only 1199 sources have counterparts in the COSMOS2020  catalog. When matching to the JWST catalog, we use the COSMOS2020  coordinates if available. If COSMOS2020  coordinates are unavailable, we use the OIR counterpart coordinates. If both are unavailable, we use the Chandra X-ray coordinates or radio coordinates. The cross-matching was done using a 0.5 arcsecond radius, and we were able to match 1228 X-ray sources with JWST counterparts in our catalog.  For the 22 sources without a JWST counterpart, we examined the JWST images and found that they are either blended sources in the ground-based image or no detections in 0.5" radius. Due to the uncertainty of the counterpart, we do not use these sources as part of our further analysis. Here, we also match our targets with the recently published COSMOS spectroscopic redshift \citep{2025arXiv250300120K} and updated the sample with additional spectroscopic redshifts and broad line classifications when available. Of the 1228 sources, 66\% have spectroscopic redshifts and for the remaining source, we adopted the photometric redshifts from \citet{2016ApJ...817...34M} which were calculated using Lephare with AGN and galaxies models and the multiband photometry in the COSMOS field.

\begin{figure}[htb]
    \centering
    \includegraphics[width=\linewidth]{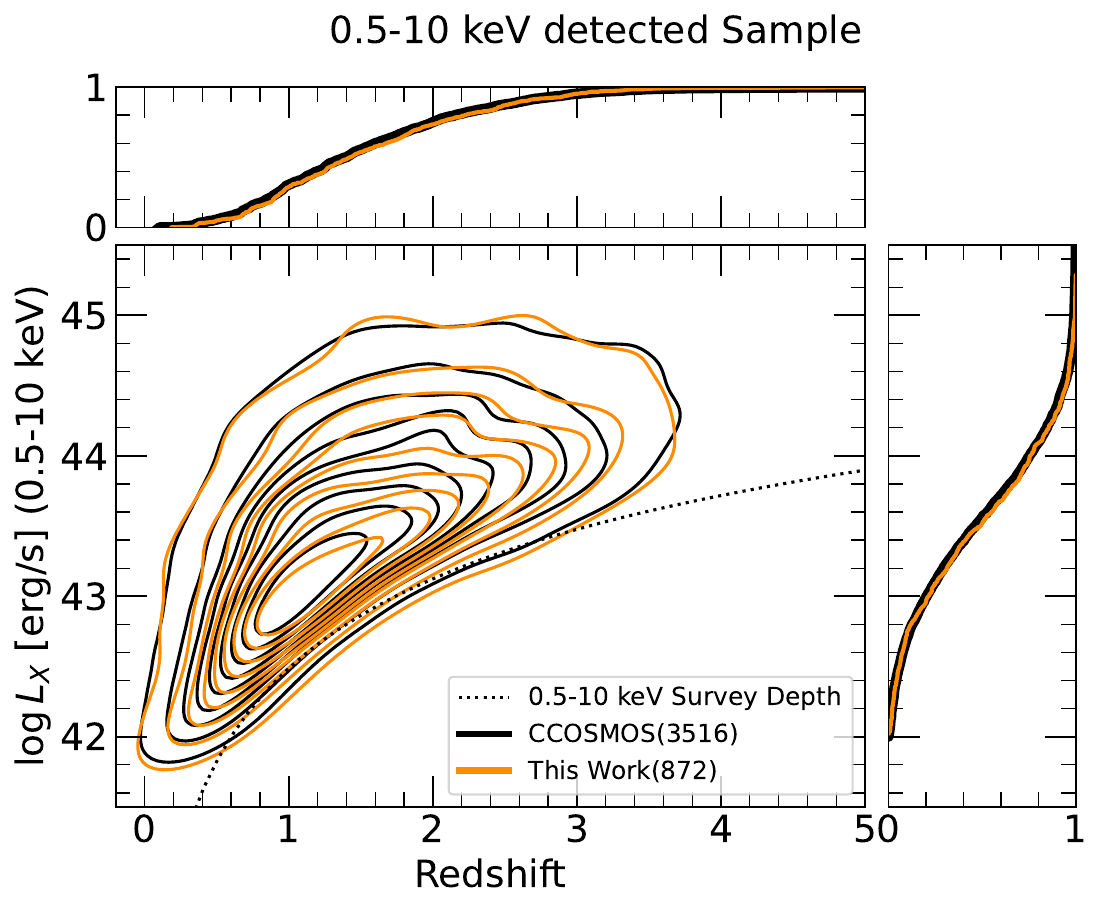}
    \caption{Redshift and 0.5-10 keV X-ray luminosity distribution of X-ray sources. The black and orange contours show the distribution of the whole CCLS catalog and our sample, respectively. The top and right panels show the normalized cumulative distribution function of the two samples. The dashed line shows the 0.5-10 keV sensitivity curve based on flux-limits from \citet{2016ApJ...819...62C}.}
    \label{fig:source_dist}
\end{figure}

For the analysis in the later sections, we limit our AGN sample to X-ray sources that satisfy the following criterion. First, they must have secure counterparts in the F444W+F277W detection image. Second, they do not lie within the HSC bright star mask of the COSMOS2020  catalog. This criterion was imposed to minimize the effect of diffraction spikes from nearby bright stars, which saturate the JWST detector. Last, they must have 0.5-10 keV X-ray luminosity above \loglx{>42}. This criterion was imposed to minimize contamination from X-ray binaries in highly star-forming galaxies, which can produce X-ray emission up to \loglx{\sim 42} \citep{2012MNRAS.419.2095M}. Among the 1228 X-ray sources, only 872 of them satisfy these criteria, while the rest were rejected due to their proximity to the bright stars. Figure~\ref{fig:source_dist} shows the X-ray luminosity and redshift distribution of the AGN sample. The majority of them are between redshift 1-3 with 2-10 keV X-ray luminosity between \loglx{=43\sim44.5}, which corresponds to moderately luminous AGN at redshift 1-3. 

\begin{deluxetable*}{lcccccccc}
\tablecaption{AGN Target coordinates, Source IDs, redshift, and magnitudes}
\label{tab:targ_coord}
\tablewidth{0pt}
\tablehead{XRID & ra\_x & dec\_x & id\_x  & z\_best\_ccls\_x & RA & Dec & mag\_F444W & magerr\_F444W \\ 
                & Deg & Deg &    &   & Deg & Deg & mag & mag } 
\startdata
XR\_09667 & 150.34101 & 2.367167 & cid\_100  & 1.59 & 150.341049 & 2.367149 & 20.442 & 0.001 \\
XR\_03897 & 149.71656 & 2.007914 & cid\_1002  & 1.623 & 149.716596 & 2.00798 & 21.165 & 0.002 \\
XR\_04756 & 149.81829 & 2.053155 & cid\_1013  & 1.231 & 149.818297 & 2.053035 & 19.954 & 0.001 \\
XR\_05393 & 149.88965 & 1.919805 & cid\_1017  & 1.329 & 149.889603 & 1.91991 & 19.341 & 0.001 \\
XR\_04397 & 149.77907 & 1.928541 & cid\_1018  & 2.421 & 149.779106 & 1.928476 & 21.558 & 0.002 \\
\enddata
\tablecomments{Table 3 is published in its entirety in the machine-readable format. A portion is shown here for guidance regarding its form and content.}
\end{deluxetable*}

\subsection{Comparison Sample Selection} \label{sec:data:compsample}

\begin{figure*}[!htb]
    \centering
    \includegraphics[width=0.48\linewidth]{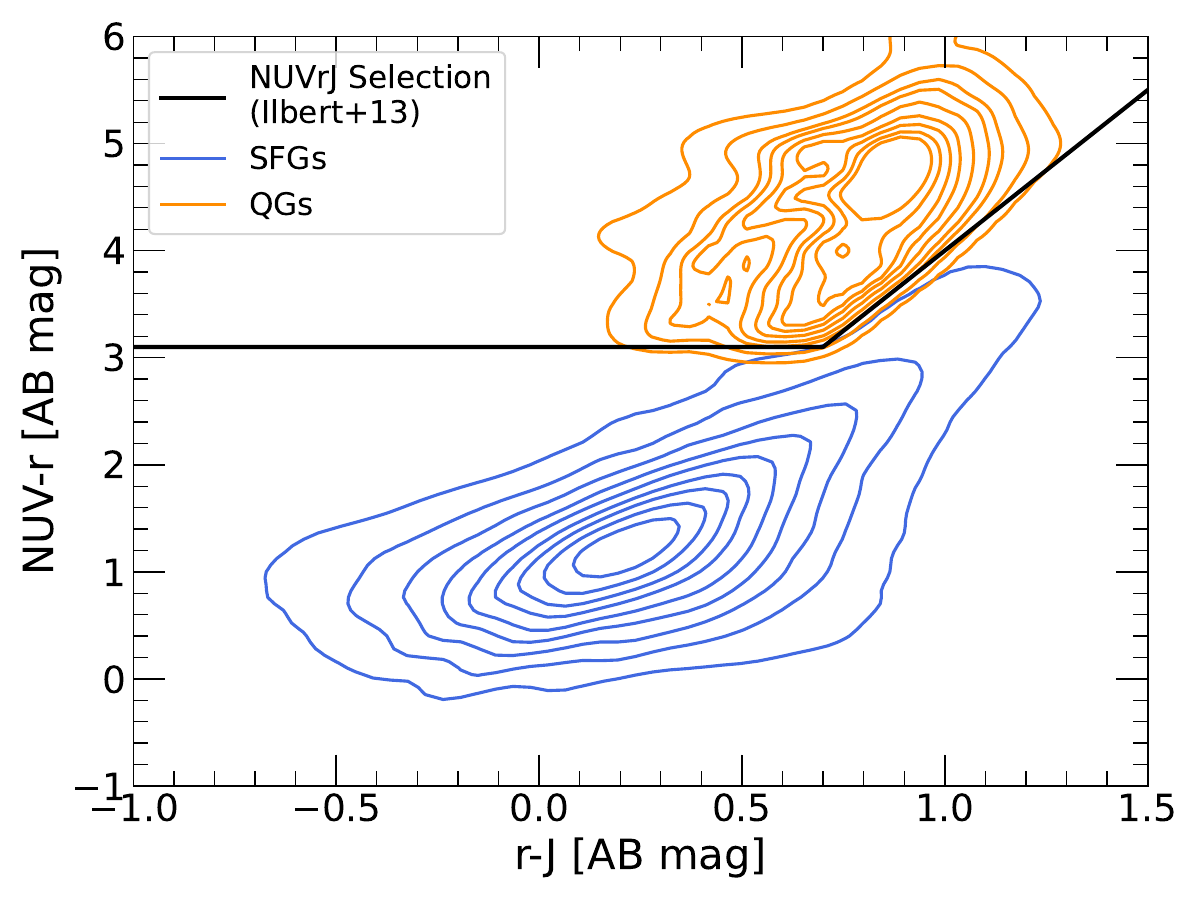}
    \includegraphics[width=0.48\linewidth]{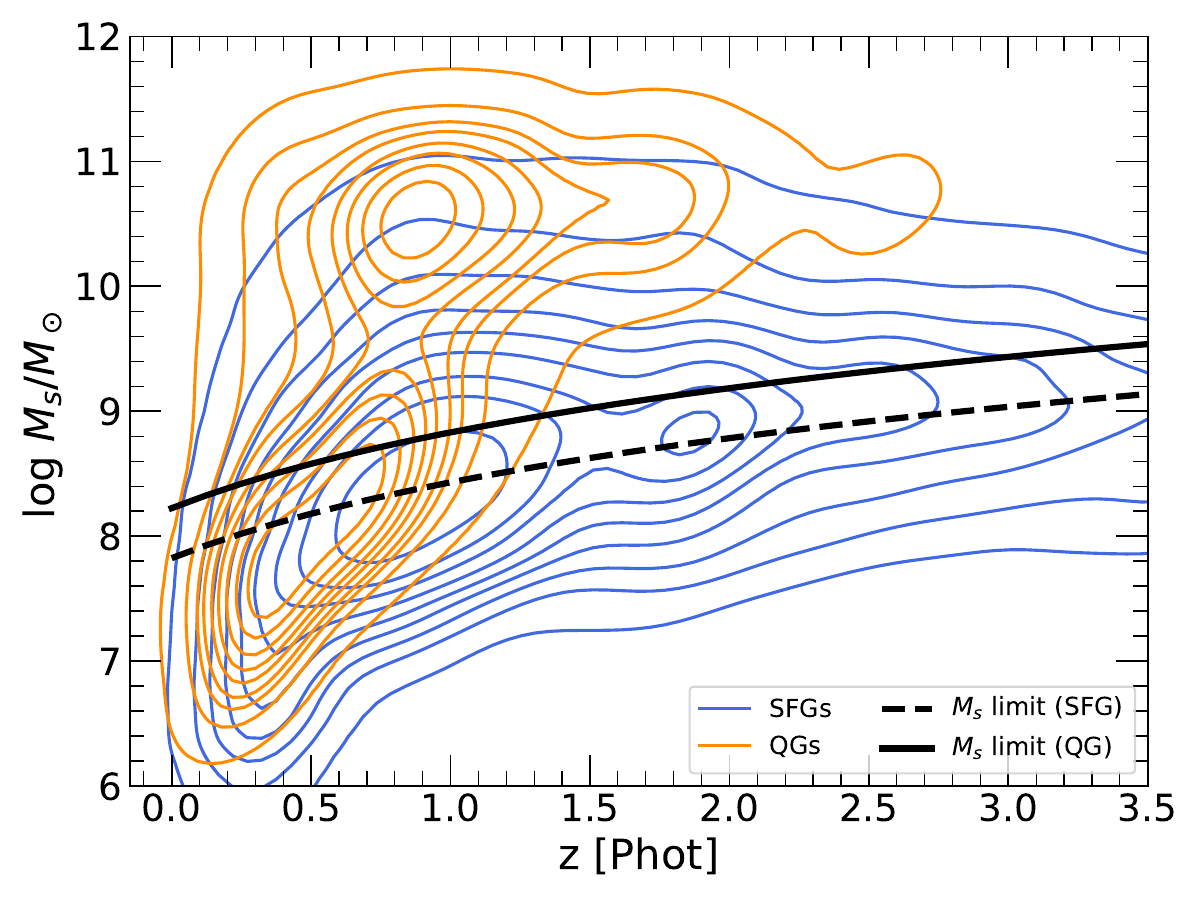}
    \caption{(Left) The NUV-r and r-J color distribution in the COSMOS2020  catalog. The blue and orange contours show the distribution of SFG and QG, respectively. (Right) The redshift and stellar mass distribution of SFG and QGs. The colors are the same as the left panel. The thick and dashed black line shows the stellar mass limit for QG and SFG, respectively. }
    \label{fig:comp_dist}
\end{figure*}

To put the size of AGN host galaxies into context with the typical galaxy population, we construct a comparison sample of star-forming (SFG) and quiescent galaxies (QG) from the COSMOS2020  catalog.  We use the NUVrJ galaxy selection \citep{2013A&A...556A..55I} to separate the galaxy sample into star-forming and quiescent galaxies. First, we chose sources best-fitted with galaxy models using the photometric redshift code LePhare \citep{1999MNRAS.310..540A,2006A&A...457..841I} and have a reduced Chi-square ($\chi^2_{\nu})<3$. Then, we limit the source to those with UltraVista $Ks$ band detection, are not in the bright star mask, and are matched with our JWST catalog within a 1" radius. 

The stellar mass of the SFGs ranges between \logMs{=7.5\sim11}, whereas QGs range between \logMs{=10\sim11.5}. \citet{2022ApJS..258...11W} performed an analysis on the mass completeness limit of the COSMOS2020 sample. They show that the completeness limit computed based on the IRAC Channel 1 depth is approximately \logMs{=8.5} at redshift 1.5 for star-forming galaxies and 0.4 dex higher for quiescent galaxies. Finally, we limit our comparison sample for further analysis to those with stellar mass above \logMs{=9.5} and have redshift between $0.2<z<3$. The redshift limit ensures that the restframe NUV band has some constraints from the deeper u-bands and that the restframe J-band is not extrapolated without constraints on the SED model from IRAC channel 2 (4.5 \um). Although the IRAC channel 3 and 4 photometry (5.6 \& 8 \um) is available, the depth is much shallower than channel 2. 

\section{Analysis} \label{sec:anal}
\subsection{2D Host Galaxy Fitting} \label{sec:anal:galfit}

We used the 2D surface brightness fitting code GALFIT \citep{2010AJ....139.2097P} to estimate the size of the AGN host galaxy. Here, we adopt a parametric 2D Sersic model to represent the AGN host galaxy. The unobscured AGN emission is modeled using a PSF model. We perform simultaneous PSF+Sersic model fitting to decompose the AGN and host galaxy components when needed.

Before we perform the surface brightness fitting, we first make cutouts of each AGN in all JWST bands. The cutout size is 3 times the segmentation size of the target in the detection image, with a minimum size of  5x5 arc-seconds. Second, we perform an additional 2D background subtraction and source detection in each band using Photutils \citep{2016ascl.soft09011B}. After subtraction of the background, the cutouts were cropped to 2 times the segmentation size of the target, and a source mask was produced by masking other sources detected in the image. Third, we set the initial conditions of the Sersic and PSF models. Photutils provides a catalog of pixel position, position angle (PA), magnitude, and axis ratio of the objects. We set the initial conditions of the Sersic components according to the source catalog given by Photutils. For model magnitudes, the Kron magnitude is assumed for the Sersic component. For the initial PSF model magnitude, we perform PSF photometry assuming a Gaussian PSF with FWHM set to that of the input PSF model.

During the fitting process, GALFIT may explore nonphysical parameter spaces, which can lead to catastrophic crashes. To prevent this, we constrain the parameter space to allow GALFIT to exit safely. We allow the centroid of the Sersic component to move freely in the source segmentation, while PSF models are allowed to move within a radius of 3 pixels from the centroid. The Sersic index is limited between 0.3 to 8. The effective radius is limited to above 0.1 pixels. Finally, the magnitude of each component was set to between 15 and 40 magnitudes. No constraints are put on the PA and the axis ratio. Objects that reach the constraints are flagged after the fitting ends. We also fit the background of the image simultaneously, but with no constraints. After running GALFIT, we estimate the uncertainties by performing a mock fitting of the best-fit model. Each mock image was constructed by adding Gaussian noise to each pixel in the best-fit image, assuming the standard deviation of the Gaussian is the same as the $1\sigma$ uncertainty from the error image. The mock fitting was performed 5 times, and the 16th and 84th percentiles of the best parameters are taken as the uncertainty. 

\subsection{Model choice} \label{sec:anal:effrad}
Since the sample is not limited to spectroscopically type-1 AGN, it is not trivial to assume the same PSF+Sersic model for all objects.  Therefore, the morphological fitting was performed using 3 sets of models: the PSF model, the Sersic model, and the PSF+Sersic model. The final results were chosen based on a combination of statistical tests using the F-test following the methods presented by \cite{2011ApJ...727L..31S} and  \citet{2009ApJ...705..639B}. We also take into account the spectroscopic classification of the AGN if available. 

First, we determine if the Sersic model is effectively indistinguishable from a point source model using an F-test. We define $$F_{a}=(\chi_{psf}^2-\chi_{Sersic}^2)/\chi_{Sersic}^2$$ where $\chi_{psf}^2$ and $\chi_{Sersic}^2$ are the best-fitted $\chi^2$ using a PSF model and Sersic model, respectively. \citet{2009ApJ...705..639B} proposed that the canonical statistical values of the F-Test are not appropriate to determine whether a source is fully dominated by AGN light (PSF model). They suggest that F-test critical values need to be determined empirically based on real unresolved point sources. Here, we choose stars from the COSMOS2020 catalog by using the BzK color selection \citep{2004ApJ...617..746D}. We further limit the star sample by choosing only those that were classified as point sources in the COSMOS ACS F814W catalogs \citep{2007ApJS..172..196K,2007ApJS..172..219L} and have magnitudes between 20-24 in the Ks-band with no additional detection within a 1 arcsecond radius.

Despite limiting the sample to F814W point sources, some contamination, overlapping, or nearby sources still linger in the star sample. To remove this contamination, we performed PSF+Sersic fitting on the star candidates, then used the Bayesian inference criterion (BIC; \cite{1978AnSta...6..461S}) difference to remove sources with contamination.  The BIC is defined as $\chi^2+ k\ln n$  where $\chi^2$ is the chi-square of the model, $k$ is the number of free parameters, and $n$ is the number of data points. The difference in the BIC ($\rm \Delta BIC$) can be used to decide which model better represents the dataset, as it takes into account the difference in the freedom between models. We identify contaminated stars by removing candidates with $\rm \Delta = BIC_{PSF} - BIC_{PSF+Sersic} \geq 100$ \citep{Kass01061995}. Typically, a difference of 10 is used, but here we used 100 based on visual identification of the contaminated star. We obtain 121, 173, 186, and 204 stars using this selection.  Using these stars, we determine the critical values as $F_{\rm a, crit}=0.01$ where sources with $F_{a}$ below this value are indistinguishable from point sources when fitted using a Sersic model. For AGNs that are indistinguishable from point sources, the effective radius in the flagged bands was removed as it can be unreliable.

Second, we determine if an additional PSF component is required for the target. In the second step, we also consider the spectroscopic classification of the AGN. If the AGN is classified as a broad-line AGN, we adopt the Sersic+PSF models unless the AGN is indistinguishable from a point source by the previous test. If the AGN has not been spectroscopically confirmed as a broad line AGN, we perform a comparison between the Sersic model and Sersic+PSF model, also using an F-test. We define $$F_{b}=\frac{\chi_{Sersic}^2-\chi_{Sersic+PSF}^2}{\chi_{Sersic+PSF
}^2}$$ where $\chi_{Sersic}^2$ and $\chi_{Sersic+PSF}^2$ are the best-fitted $\chi^2$ using a Sersic model and Sersic+PSF model, respectively. The critical value of the classification is determined based on the spectroscopically confirmed broad-line AGN that passes the first test and visual identification of the image and residuals. We set the critical values as $F_{\rm b, crit}=0.05$. We also perform a check using $\rm \Delta BIC= BIC_{Sersic} - BIC_{Sersic+PSF}$ and confirmed that the broad-line AGN have $\rm \Delta BIC >10$ \citep{2025ApJ...979..215T,Kass01061995}. Table~\ref{tab:fitstats} summarizes the number of AGN fitted with a PSF, PSF+Sersic, Sersic model, and failed fitting. 

\begin{deluxetable}{lcccc}[!htb]
    \tablecaption{Number of AGN in each Model Case.}
    \label{tab:fitstats}
    \tablehead{
        \colhead{Band} & \colhead{PSF} & \colhead{PSF+Sersic} & \colhead{Sersic} & \colhead{Failed }}
        \startdata
        \multicolumn{5}{c}{Broad line AGN (213)} \\
        \hline
        F090W & 6 & 6 & 0 & 0 \\
        F115W & 56 & 155 & 0 & 2 \\
        F150W & 35 & 177 & 0 & 1 \\
        F200W & 1 & 9 & 0 & 2 \\
        F277W & 4 & 209 & 0 & 0 \\
        F356W & 0 & 12 & 0 & 0 \\
        F410M & 0 & 12 & 0 & 0 \\
        F444W & 4 & 208 & 0 & 1 \\
        \hline 
        \multicolumn{5}{c}{Non-Broad-line AGN (659)} \\
        \hline
        F090W & 13 & 3 & 31 & 3 \\
        F115W & 61 & 26 & 536 & 36 \\
        F150W & 19 & 83 & 548 & 9 \\
        F200W & 1 & 13 & 34 & 2 \\
        F277W & 4 & 262 & 390 & 3 \\
        F356W & 1 & 17 & 30 & 2 \\
        F410M & 1 & 14 & 32 & 3 \\
        F444W & 3 & 290 & 360 & 6 
\enddata
\end{deluxetable}

\subsection{Estimation of host galaxy stellar mass} \label{sec:anal:mstar}

\begin{figure}[h!]
    \centering
    \includegraphics[width=\linewidth]{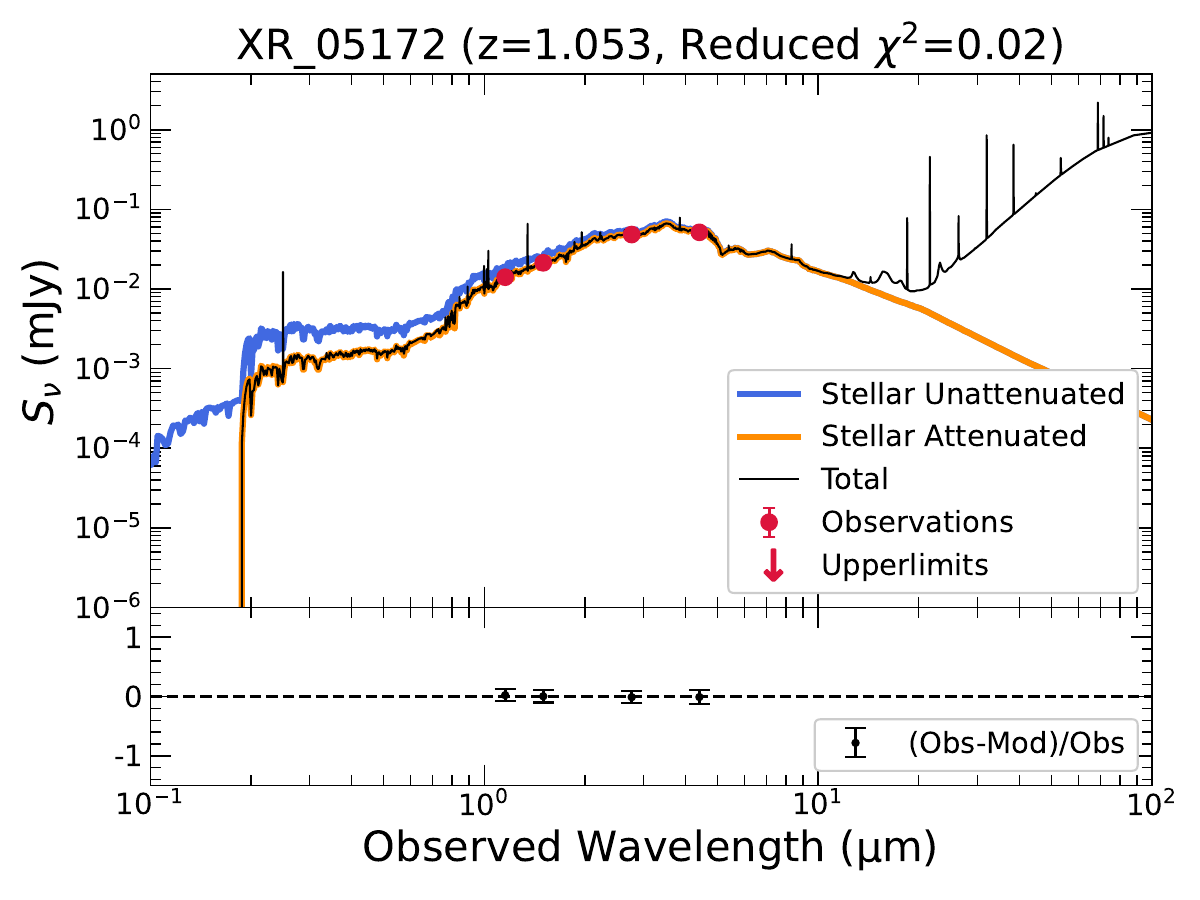}
    \caption{An example of the decomposed SED fitting using CIGALE. The red dots show the JWST NIRCam detection. The blue and orange lines show the best-fitted unattenuated and attenuated stellar components. The total SED model is shown with black lines. The bottom panel shows the normalized error of the best-fitted SED against the observed data as a function of wavelength.}
    \label{fig:sedfit}
\end{figure}

We perform spectral energy distribution (SED) fitting using CIGALE \citep{2019A&A...622A.103B,2022ApJ...927..192Y} to estimate the stellar mass of the AGN host galaxy. We limit the SED fitting analysis to AGN host galaxies with fluxes in all 4 NIRCam bands from the COSMOS-Web Survey, but we also use the additional 4 bands for 62 AGN that are also in the PRIMER survey area. For each AGN, we remove the AGN contribution from the photometry in each band by subtracting the flux estimated from the best-fitted PSF component, which represents the AGN emission. For the SED fitting models, we assume an exponentially declining star formation history and a Chabrier initial mass function \citep{2003PASP..115..763C}. We assume a modified starburst extinction law with the color excess as a free parameter. This model is based on the starburst attenuation curve presented by \citet{2000ApJ...533..682C} with extensions added from \citet{2002ApJS..140..303L} in the far-ultraviolet.  Since we have F444W, which overlaps with hot interstellar dust and PAH emission in lower redshift galaxies, we adopted the 2-parameter dust models from \citet{2014ApJ...784...83D}. We chose this model due to its simplicity, given that we do not incorporate constraints on the dust continuum. We do not fit with the AGN components, assuming that the AGN contribution has been removed from the total flux. The fitting settings are shown in Table~\ref{tab:sedparam}. We adopt an additional 10\% systematic uncertainty for each flux in each band. As shown by recent efforts by the PEARLS survey \citep{2024PASP..136b4501M}, the absolute flux calibration uncertainty of NIRCam may have an uncertainty of up to 0.05 magnitudes in the current calibration sets. Figure~\ref{fig:sedfit} shows an example of our best-fitted SED. Based on the best-fit SED, $\sim93\%$ of the AGN sample would be classified as star-forming galaxies based on the NUVrj colors.  This is consistent with previous work on the colors of AGN host galaxies shows that most X-ray selected AGN have colors consistent with SFGs \citep{2008ApJ...675.1025S,2014MNRAS.440..339G,2023RNAAS...7..165C,2024MNRAS.529.3610V}. However, we would like to note that we have limited ability to constrain the UV portion of the SED due to a lack of photometric bands with sufficient angular resolution. Therefore, while we consider our sample to be similar to SFGs, the color classification should be taken at face value. The AGN host stellar masses are given in table \ref{tab:targprop} For our comparison sample, we used the stellar mass in the COSMOS2020 catalog estimated with Lephare, which also uses the Chabrier IMF.

\begin{table*}[ht]
	\centering
	\caption{CIGALE Parameters Settings}
	\label{tab:sedparam}
	\begin{tabular}{lc} 
	    \hline
        \multicolumn{2}{c}{Star-formation History (sfhdelayed)}  \\
        \hline
            $e$-folding time of main population ($\tau_{\rm Main}$) & 100,200,400,600,800,1000,2000,4000,6000,8000 Myr \\
            Age of the main stellar population ($\rm Age_{\rm Main}$) & 500,1000,1200,1400,1600,1800,2000 \\ 
                                                                      & 3000,4000,6000,8000,10000 Myr \\ 
            $e$-folding time of the late starburst population ($\tau_{\rm Burst}$) & 3 \\
            Age of the late burst in Myr ($\rm Age_{\rm Burst}$) & 100.0 \\
            Mass fraction of the late burst population ($\rm f_{\rm Burst}$) & 0.0,0.01 \\ 		
        \hline
        \multicolumn{2}{c}{SSP Model (BC03)}  \\
        \hline
            Initial mass function (IMF) & Chabrier \\ 
            Metallicity & 0.02, 0.05 \\
            Old/Young Stellar Age Separation & 10.0 Myr \\
        \hline
        \multicolumn{2}{c}{Dust Attenuation (dustatt\_modified\_starburst)}  \\
        \hline
            The colour excess of the nebular lines (E(B-V)) &  0.05,0.1,0.2,0.4,0.6,0.8,1,1.2,1.4,1.6\\
            Stellar E(B-V) conversion factor & 0.44 \\
            Ext law emission lines & \citet{1989ApJ...345..245C} \\
            Ratio of total to selective extinction ($\rm R_V$) & 3.1 \\
        \hline
        \multicolumn{2}{c}{Dust Emission \citep{2014ApJ...784...83D}}  \\
        \hline
            $\rm f_{AGN}$ & 0 \\
            Alpha slope ($\alpha$) & 0.0625,1.0,2.0,3.0,4.0 \\
        \hline
        \hline	\end{tabular}
\end{table*}

\subsection{Estimation of Restframe 1\,\um Effective Radius and Sersic Index} \label{sec:anal:resteffrad}

The effective radius ($R_e$) of a 2D Sersic profile is defined as the length along the semi-major axis that contains half the total light. Since it depends on the rest-frame wavelength, we correct the observed effective radius using the method presented in  \citet{2014ApJ...788...28V}. Equation~\ref{eq:effrac} shows the correction of the rest-frame effective radius from the observed effective radius,

\begin{equation}
    R_{\rm e}=R_{\rm e, pivot}\bigg( \frac{1+z}{1+z_{p}} \bigg)^{\frac{\Delta \log R_{\rm e}}{\Delta \log \lambda}}
    \label{eq:effrac}
\end{equation}

where $R_{\rm e, pivot}$ is the pivot effective radius in kilo parsec, $z$ is the redshift, $z_p$ is the pivot redshift, and $\frac{\Delta \log R_{\rm e}}{\Delta \log \lambda}$ is the color-size gradient term. The pivot redshift is the redshift at which the effective radius of interest falls into the pivot band.  We choose the 1 \um effective radius because the size gradient and the effective radius can be estimated with two JWST bands covering shorter and longer wavelengths of 1 \um. We use the 4 bands covered by COSMOS-Web as our pivot sizes in bins of redshift according to Table~\ref{tab:pivotsize} and also use the Sersic index measured in the pivot band for our later analysis.

\begin{deluxetable}{lclc}[!htb]
\tablecaption{Pivot Redshift and Filter}
\label{tab:pivotsize}
\tablehead{
    \colhead{[$z$, $z_p$]} & \colhead{Pivot Filter}  }
    \startdata
    0.00, 0.15 & F115W \\
    0.15, 0.50 & F150W \\
    0.50, 1.77 & F277W \\
    1.77, 3.40 & F444W 
\enddata
\end{deluxetable}

The size gradient ($SG=\Delta \log R_{\rm e}/ \Delta \log \lambda $) is the average change in the effective radius as a function of wavelength. \citet{2014ApJ...788...28V} showed it depends on stellar mass, redshift, and population. For galaxies between redshifts 0 to 0.5, 0.5 to 1.77, and 1.77 to 3.44, we use F115W and F150W, F150W and F277W, and F277W and F444W to calculate the size gradient for both the AGN and comparison galaxy sample, respectively.  We only use galaxies whose effective radius or the Sersic index is not an upper limit to estimate the size gradient. We also calculated the size gradient using 8 bands for galaxies in the PRIMER region and compared it with the 4-band cases as a consistency check. We find that the determination of the size gradient agrees well with each other. We use the F115W, F150W, F277W, and F444W for consistency over the whole sample for our size gradient and effective radius measurements.

\begin{figure}[htb!]
    \centering
    \includegraphics[width=\linewidth]{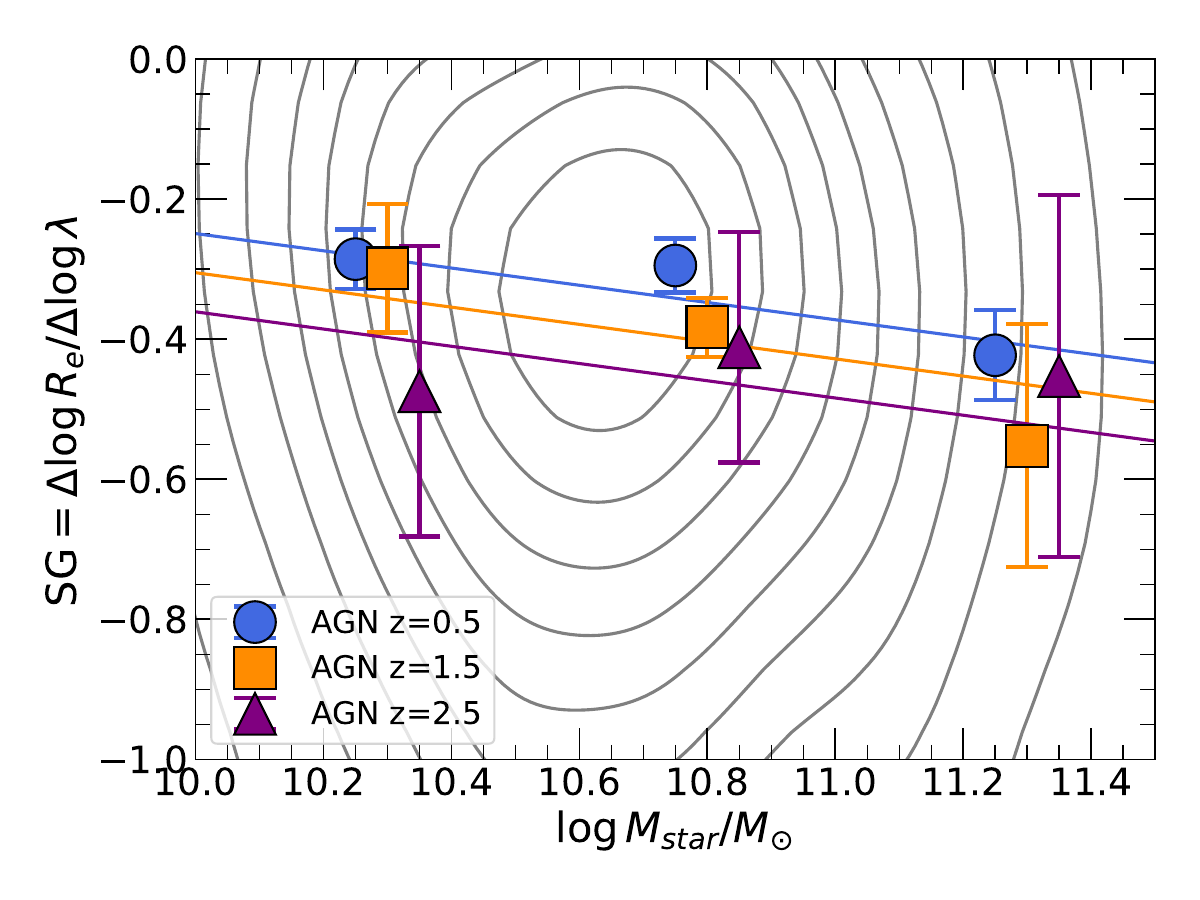}
    \caption{Size gradient of AGN host galaxies. The contours show the distribution in the size gradient of AGN host galaxies. Markers show the median size gradient in bins of redshift and stellar mass. For clarity, the markers are offset to the right by 0.02 dex in redshift bins from the lowest redshift. The best-fit size gradient function is shown with colored lines.}
    \label{fig:cgrad1um}
\end{figure}

For the AGN host galaxy, we first examine the 1 \um size gradient in bins of stellar mass and redshift. We observe a mild redshift and stellar mass dependence in the size gradient when the sample is binned in figure~\ref{fig:cgrad1um}. We fitted the size gradient in bins of stellar mass and redshift, assuming that the size gradient depends linearly on redshift and stellar mass. The best-fitted functional form of the size gradient is

\begin{equation}
    \begin{split}
    {\rm SG} = & -0.31(\pm0.05) -0.12(\pm0.05)\log \bigg( \frac{M_s}{5\times 10^{10}}\bigg) 
    \\ &  -0.06(\pm0.03) z    
    \end{split}
\end{equation}

where $z$ is the redshift and $M_s$ is the galaxy stellar mass. The restframe $\rm 1 \mu m$ effective radius, Sersic indexes, and axis ratios are given in table \ref{tab:targprop}

\begin{figure*}[ht]
    \centering
    \includegraphics[width=\linewidth]{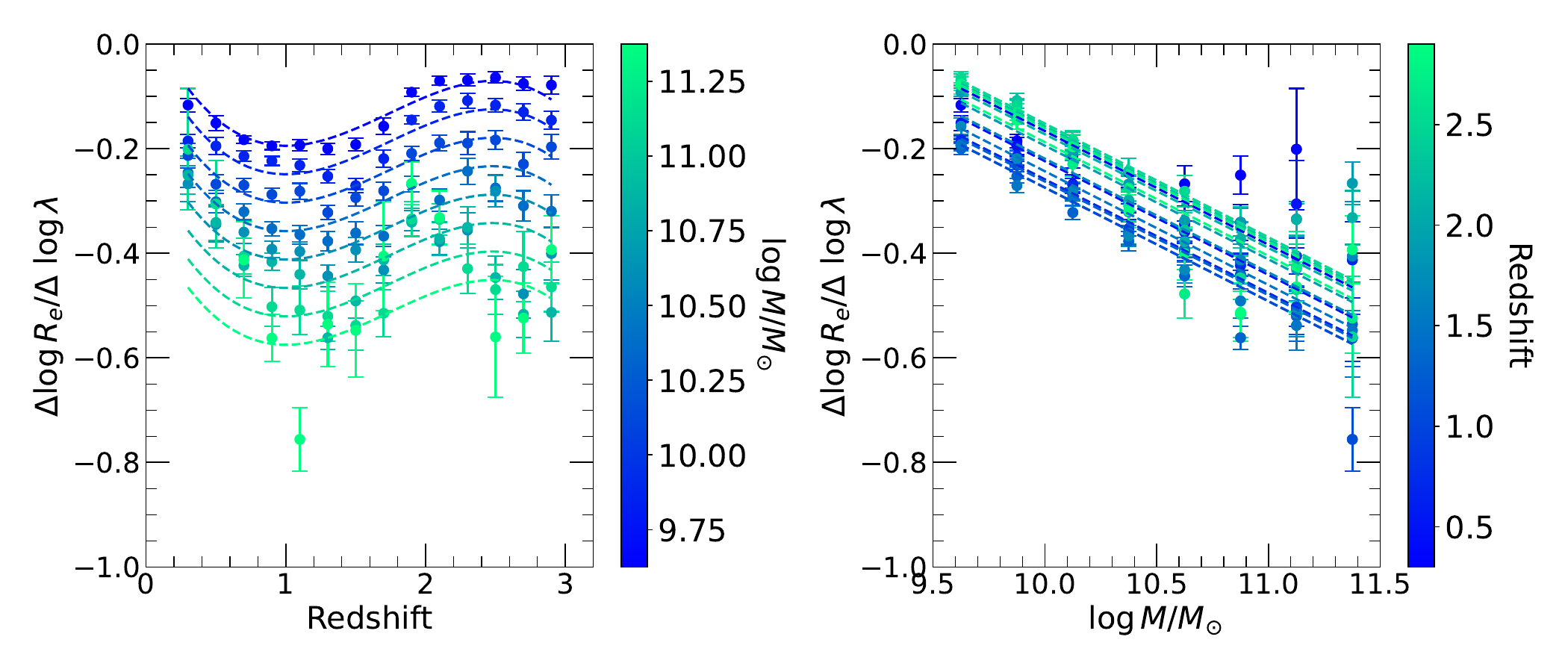}
    \includegraphics[width=\linewidth]{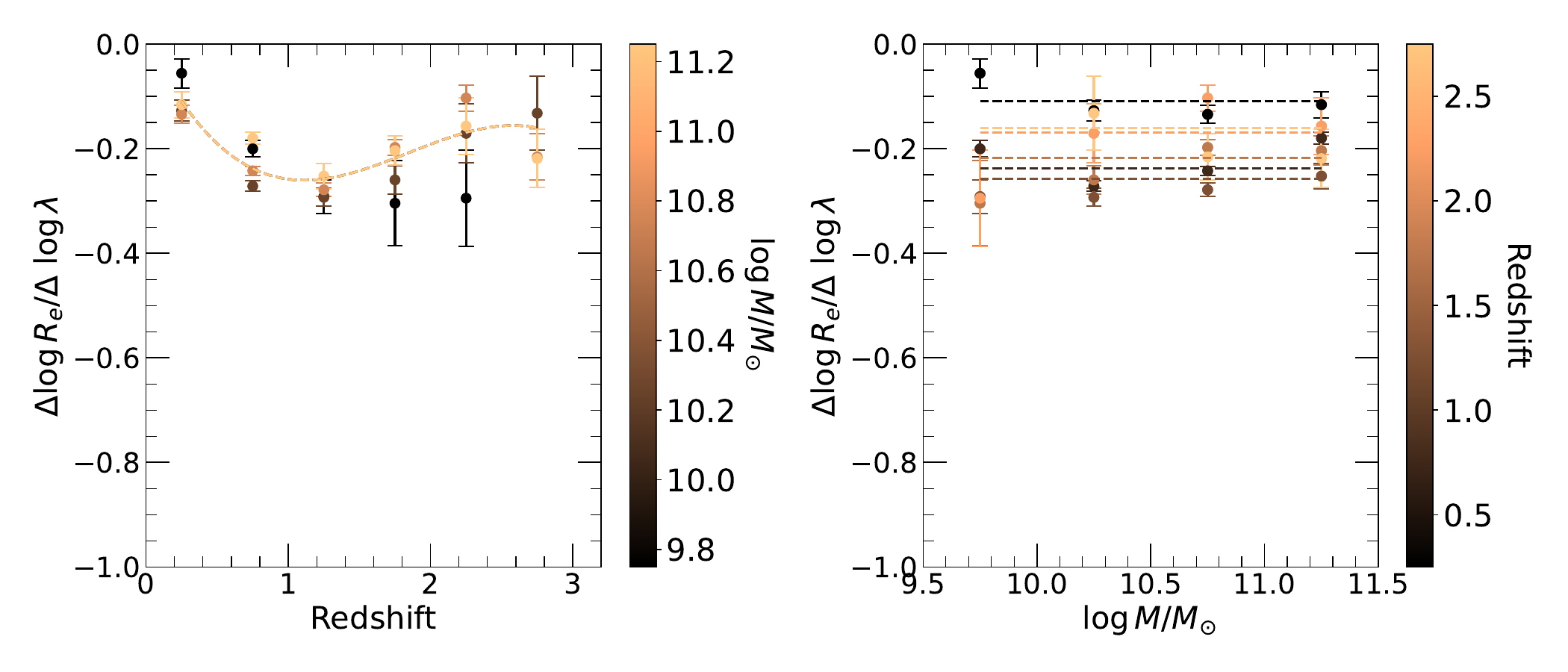}
    \caption{The binned size gradient in redshift (left column) and stellar mass (right column) of SFG (top panels) and QG (bottom panel). The best-fitted functional form is shown as the dashed colored lines.  }
    \label{fig:cgrad_type}
\end{figure*}

We perform a similar analysis for the SFG and QG as AGN host galaxies. We calculate the median size gradient and standard uncertainty of the mean in bins of redshift and stellar mass with a requirement that there are at least 10 galaxies in each bin. The median size gradient of star-forming galaxies and quiescent galaxies is shown in figure~\ref{fig:cgrad_type}. For SFG, we see a clear decreasing trend with increasing stellar mass. This decline is also seen in previous work by \citet{2014ApJ...788...28V}. On the other hand, quiescent galaxies do not show an obvious trend with stellar mass, similar to previous work by \citep{2011ApJ...735...18G,2012MNRAS.421.1007K}.  In redshift, the size gradient of both populations shows a decline between redshifts 0 to 1 and then an increase towards redshift 3. To account for both the observed redshift and stellar mass dependence of the size gradient. We fit the size gradient assuming a 3rd degree polynomial with a linear dependence in the stellar mass. The function is described as 

\begin{equation}
    \begin{split}
    {\rm SG}(z,M_s) = &  c_0 + c_1z + c_2z^2 + c_3z^3 + \\ 
    & c_4\log \bigg(\frac{M_s}{5 \times 10^{10} M_\odot}\bigg)   
    \end{split}
\end{equation}

where $z$ is the redshift and $M_s$ is the galaxy stellar mass. Since the quiescent galaxy sample does not show a clear dependency on the stellar mass, we fit the size gradient without the stellar mass dependence terms.  The best-fitted coefficients are shown in Table~\ref {tab:sgrad_coeff}.  The size gradient of AGN host galaxies is similar to SFGs of the same stellar mass and smaller than QGs. At face value, the size gradient shows a milder dependence on stellar mass than SFG, similar to QGs. Finally, we use the functional form of the size gradient to estimate the restframe 1\,$\micron$ effective radius following equation~\ref{eq:effrac}. Figure~\ref{fig:re_ns_dist} shows the distribution of AGN, SFG, and QG in the $\log R_e$ and Sersic index plane.

\begin{deluxetable*}{lCCCCC}[!htb]
\tablecaption{Size Gradient Function coefficients of SFG and QGs}
\label{tab:sgrad_coeff}
\tablehead{
    \colhead{Sample} & \colhead{$c_0$} & \colhead{$c_1$} & \colhead{$c_2$}& \colhead{$c_3$} & \colhead{$c_4$}  }
    \startdata
    SFG & -0.181 \pm 0.014 & -0.570 \pm 0.032 & 0.400 \pm 0.022 & -0.077 \pm 0.005 & -0.217 \pm 0.004 \\
    QG  &  0.012 \pm 0.021 & -0.572 \pm 0.060 & 0.368 \pm 0.050 & -0.067 \pm 0.012 & -
    \enddata
\end{deluxetable*}

\subsection{Non-parametric Morphology} \label{sec:anal:npar}

We also calculated Gini ($G$) and  $M_{20}$ non-parametric parameters for our comparison galaxies and AGN host galaxies. The Gini coefficient is a measure of the spread of light in the pixel distribution, while the  $M_{20}$ parameter is the normalized second-order moment of the brightest 20\% pixels \citep{2003ApJ...588..218A,2004AJ....128..163L}.  We focus mainly on the Gini and $M_{20}$ parameters, which are useful for classifying galaxies into mergers and non-mergers \citep{2004AJ....128..163L,2008ApJ...672..177L}. As shown in \citet{2008ApJ...672..177L}, merging galaxies occupy the region above $ G > -0.14 M_{20} +0.33$. It can also distinguish between spiral galaxies and spheroidal galaxies. For galaxies below the merger boundary previously described, galaxies above $G > 0.14 M_{20} +0.80$ are spheroidal in shape, whereas those below are spirals. We perform the calculations using Statmorph \citep{2019MNRAS.483.4140R} for both the comparison galaxies and AGN hosts using the same segmentation and masking of other objects. If the AGN requires a second PSF component, we perform the non-parametric calculations on the PSF-subtracted images. We adopted the pivot bands listed in table~\ref{tab:pivotsize} for the 1 \um non-parametric parameters. The non-parametric morphologies are given in table \ref{tab:targprop}.

\begin{deluxetable*}{lccccccccc}
\tablecaption{AGN Host Stellar Mass and Rest-frame 1$\rm \mu m$ Shapes Parameter}
\tablehead{XRID & $\log (M_s/M_{\odot})$ & $\sigma_{\log (Ms/M_{\odot})}$ & $R_{e, 1\mu m}$ & $R_{\rm e, Band}$ & $n_{\rm Sersic, 1\mu m}$ & $\rm 1\mu m$ Axis ratio & GINI & M20 & ASYM \\  & dex & dex & kpc & kpc &  &  &  &  &  }
\label{tab:targprop}
\tablewidth{0pt}
\startdata
XR\_03575 & 10.29 & 0.16 & 2.37 & F277W & 2.42 & 0.88 & 0.555 & -1.701 & 0.106 \\
XR\_03735 & 10.55 & 0.11 & 2.07 & F444W & 2.85 & 0.34 & 0.532 & -1.886 & 0.055 \\
XR\_03887 & 10.53 & 0.21 & 0.3 & F444W & 2.98 & 0.53 & 0.418 & -1.529 & 0.082 \\
XR\_03897 & 10.21 & 0.12 & 2.15 & F277W & 1.82 & 0.92 & 0.59 & -2.048 & 0.112 \\
XR\_03902 & 10.54 & 0.18 & 2.17 & F277W & 7.86 & 0.5 & 0.631 & -2.342 & 0.101 \\
\enddata
\tablecomments{Table 8 is published in its entirety in the machine-readable format. A portion is shown here for guidance regarding its form and content.}

\end{deluxetable*}

\section{Results} \label{sec:results}

\subsection{1 \um Effective radius, Sersic Index, \& Axis Ratio Distributions} \label{sec:res:Sersic}

\begin{figure}[!ht]
    \centering
    \includegraphics[width=\linewidth]{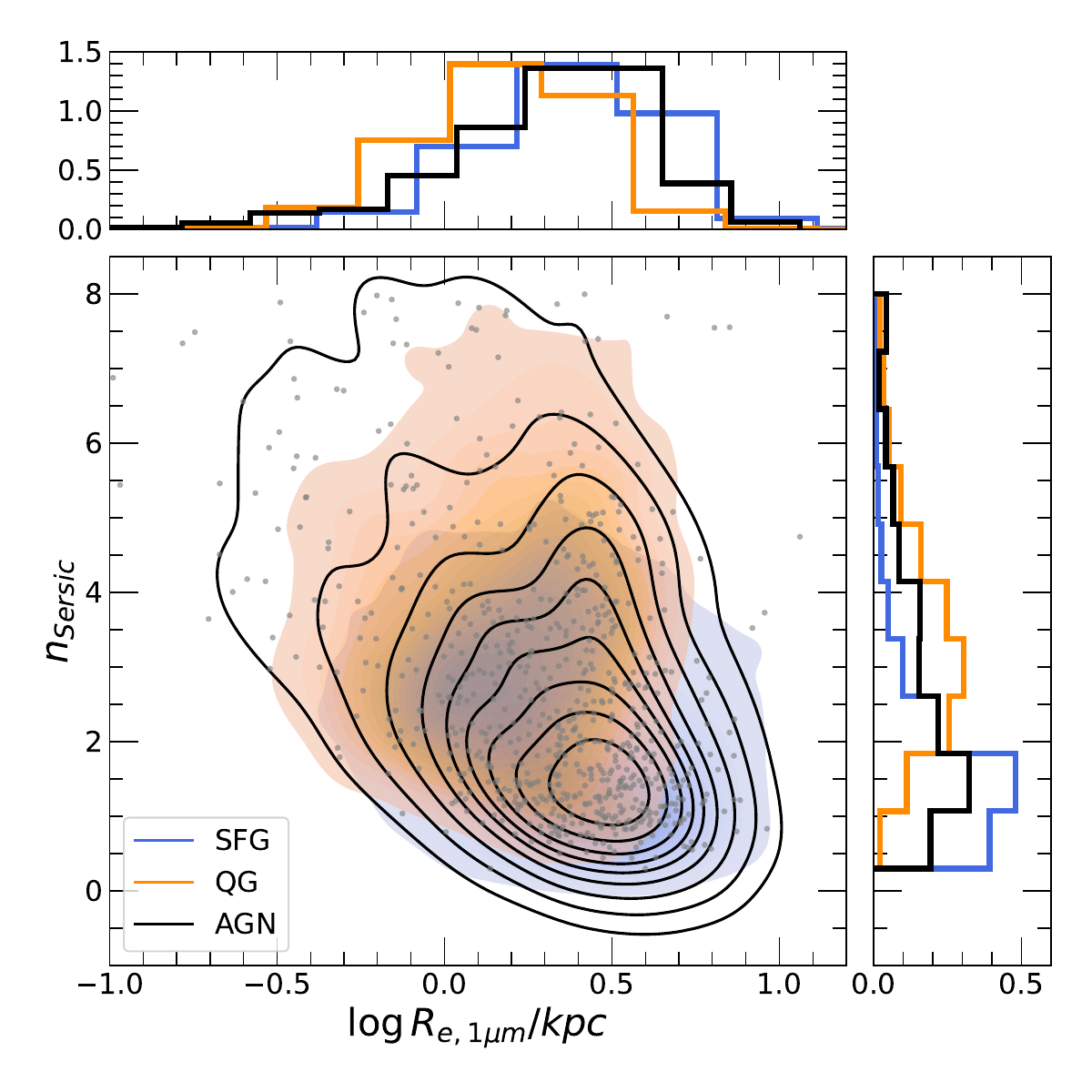}
    \caption{Effective radius and Sersic index distribution. AGN host galaxies, SFG, and QG are shown in black, blue, and orange contours. Gray dots show individual AGN The top and right panel shows the $\log R_{\rm eff}$ and Sersic index distribution of the 3 samples. }
    \label{fig:re_ns_dist}
\end{figure}

The Sersic index ($n_{\rm s}$) describes the surface brightness distribution. In the local universe, disk-like galaxies have a Sersic index of 0.5-2, whereas spheroidal galaxies (or bulge-dominated galaxies) tend to have a large Sersic index $n_{\rm s}\geq 2$. It is used as a first-order estimate of the galaxy shape. Figure~\ref{fig:re_ns_dist} shows the 1\um effective radius and Sersic index distribution of SFG, QG, and AGN host galaxies. Most galaxies have sizes between 1 to 10 kpc, with QG samples smaller than SFG galaxies. AGN host galaxies have a much broader distribution, overlapping with both SFG and QG samples. The Sersic index of SFG and QG shows a different distribution, with QG galaxies having a median Sersic index of $\sim 3$ while the majority of the former have a Sersic index $\sim 1$. Similar to the effective radius, AGN host galaxies have a broad Sersic index distribution. 

Compared with previous work, we find a slightly larger Sersic index in AGN host galaxies than previous studies, which estimate Sersic indexes of $\sim 1-3$ at 0.5 \um wavelengths \citep{2015A&A...573A..85R,2021ApJ...918...22L,2024MNRAS.527.4690L,2024ApJ...962..139Z,2024ApJ...962...93Z}. The difference in our results with previous work could be due to the wavelength of interest, where previous work utilized rest-frame optical data and UV sizes (eg. i-band).  We utilize near-infrared observations (mostly F277W observations), which are more sensitive to older stellar populations and less sensitive to dust attenuation. 

It is worth mentioning that the shape of galaxies can have a dependence on the stellar mass of the galaxies themselves. Therefore, it is wise to match the stellar mass and redshift distribution of the galaxies to the same distribution before conducting a statistical test between the shape distributions (eg. Sersic index, axis ratio, Gini-$M_{20}$). To do this, we perform a 1-1 propensity score matching to match the SFG and QG stellar mass and redshift distribution. This process constructs a matched redshift and stellar mass SFG and QG subsample by drawing 1 SFG and QG that has a similar stellar mass and redshift for each AGN. In principle, one can use the closest SFG and QG as a comparison or an optimal weighing scheme that minimizes the tolerance in the sample. Instead, we identify counterpart SFG and QG that have similar redshift and stellar mass within 0.1 tolerance in redshift and 0.1 dex tolerance in stellar mass of the AGN host galaxy. This is based on the expected uncertainty in stellar mass and redshift and allows us to bootstrap different matched SFGs and QGs to estimate uncertainties. Typically, each AGN host galaxies have $>20$ counterpart SFG and QG, available for sampling.  

To evaluate the uncertainties in the AGN distribution, we also constructed mock AGN samples by bootstrapping the full AGN sample. Each mock sample has the same number of AGN as the total sample. We construct several mock QG and SFG sub-samples for each mock AGN sample by applying the previously described propensity score matching. We used this method to compare the Sersic index distribution and the axis ratio distribution of the AGN host galaxies. After matching the stellar mass and redshift, the AGN host shape distribution is compared against SFG and QG using a Kolmogorov–Smirnov test (KS-test).

To compare the Sersic ratio distributions, we first bin the sample in bins of redshift from 0 to 3 with and stellar mass between $\log M_s=10-11.5$ with bin sizes of 0.5 dex. This mass range corresponds to the range in which most AGN in our sample fall. We perform the redshift and stellar mass matching previously described in each bin. We first draw 100 mock AGN samples. For each mock AGN sample, we draw  10 SFG and QG samples for each mock AGN. In total, 1000 mock samples of SFG and QG are generated for each bin.

\begin{figure*}[!ht]
    \centering
    \includegraphics[width=\linewidth]{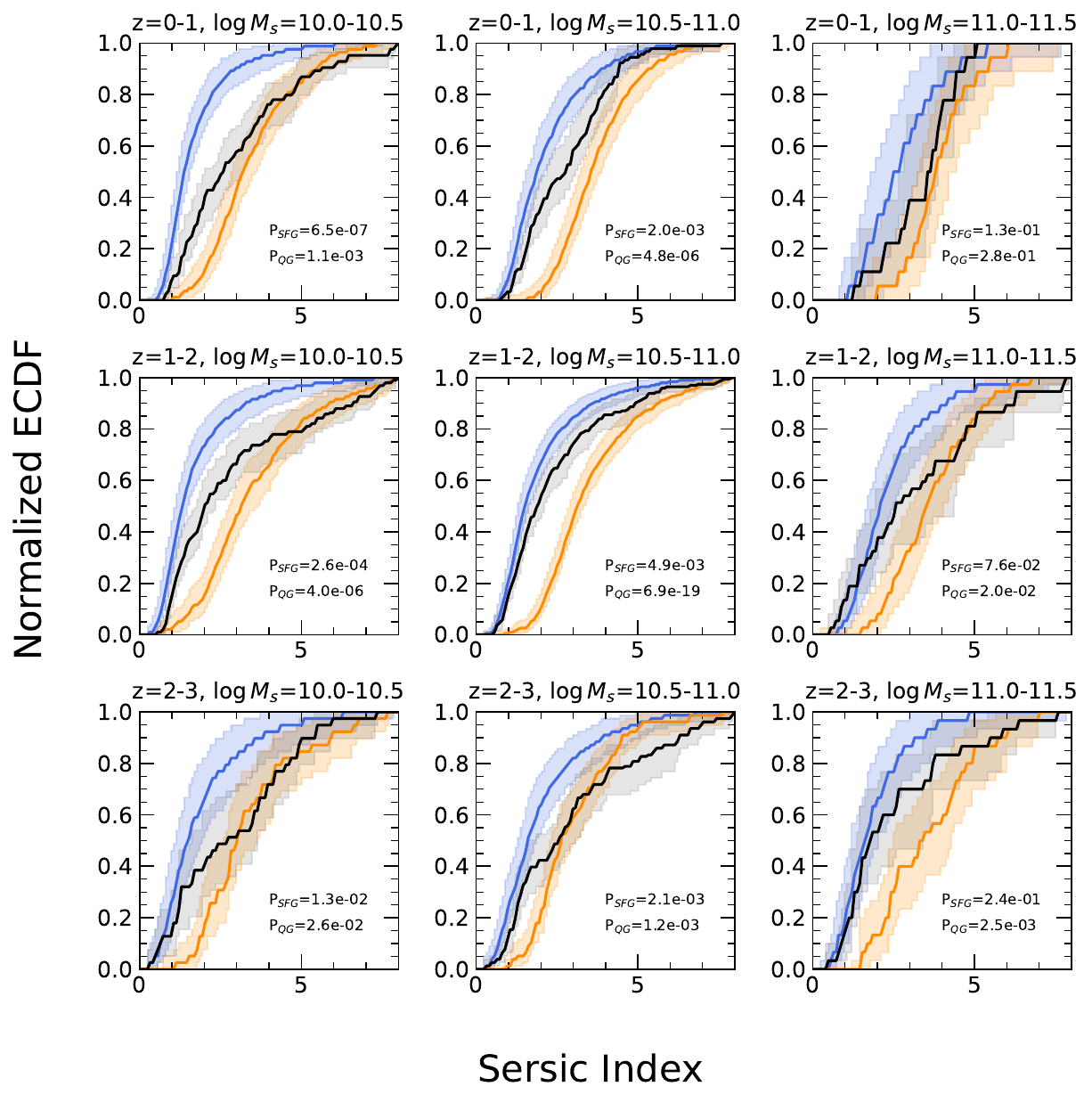}
    \caption{The normalized cumulative distribution of the Sersic index binned in redshift and stellar mass. The distribution of AGN host galaxies, SFG, and QG is shown in black, blue, and orange, respectively. The shaded region shows the 16th and 84th percentiles of the distribution.}
    \label{fig:Sersic_dist}
\end{figure*}

\begin{figure*}
    \centering
    \includegraphics[width=\linewidth]{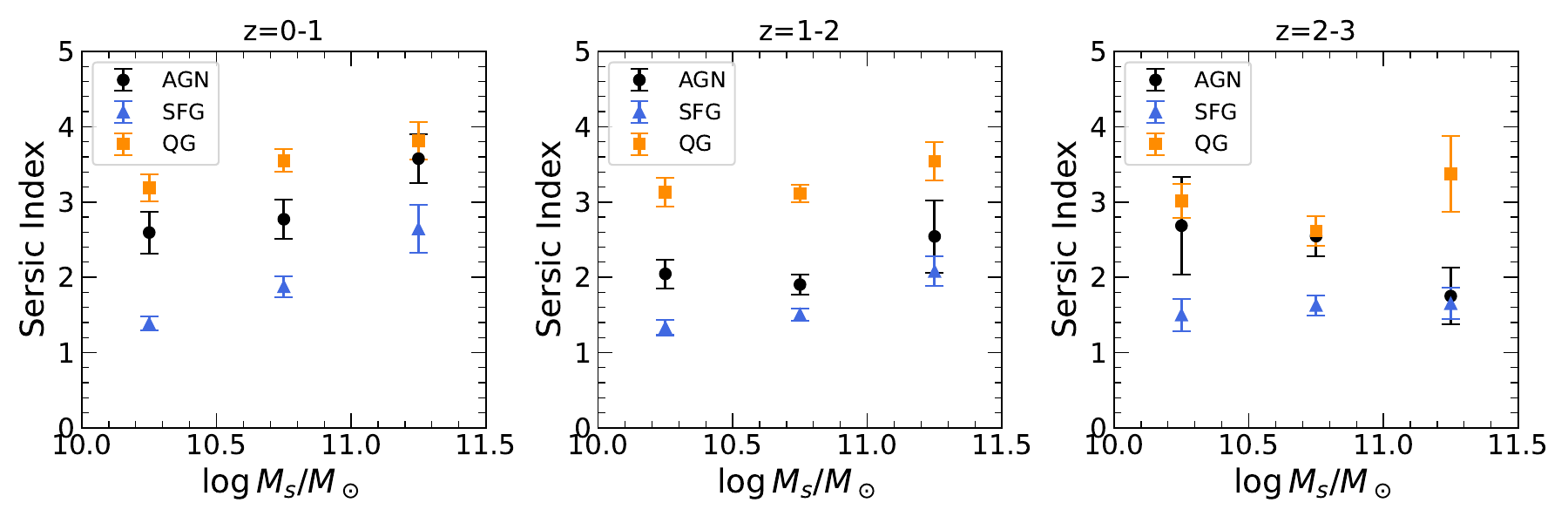}
    \caption{The median Sersic index of AGN host galaxies, SFG, and QGs shown as black round symbols, blue triangles, and orange squares in redshift and stellar mass bins.}
    \label{fig:binned_serc}
\end{figure*}

Each bin is shown in figure~\ref{fig:Sersic_dist}, we compute the KS statistics and p-value comparing the Sersic index distribution of AGN against SFG and QG, respectively.  We find that in most stellar mass \& redshift bins, we reject the null hypothesis in which AGN and SFG, as well as AGN and QG, have the same Sersic index distribution at a 99\% confidence. This suggests that the distribution of shapes of AGN host galaxies differs from SFG and QG.   We also compare the median Sersic index of the AGN host galaxies against SFG and QG as a function of stellar mass in 3 redshift bins, shown in figure~\ref{fig:binned_serc} after matching the stellar mass distributions in each redshift bin. Overall, the Sersic indices increase with stellar mass for all types, except in redshift 2-3, where the Sersic index has a weaker dependence on stellar mass. In all redshift bins, QG shows the largest Sersic index, followed by AGN, and lastly SFG. 

Besides the Sersic index and the effective radius, GALFIT also provides the projected axis ratio between the semi-minor and semi-major axes of the 2D Sersic profile. On an individual object basis, the projected axis ratio is not so powerful due to inclination effects. On a population level, the projected axis ratio can be used as a crude classifier of morphology with QGs on average rounder than SFG \citep{2013ApJ...778..115P,2014ApJ...792L...6V,2019ApJ...871...76H}. 

\begin{figure}
    \centering
    \includegraphics[width=0.8\linewidth]{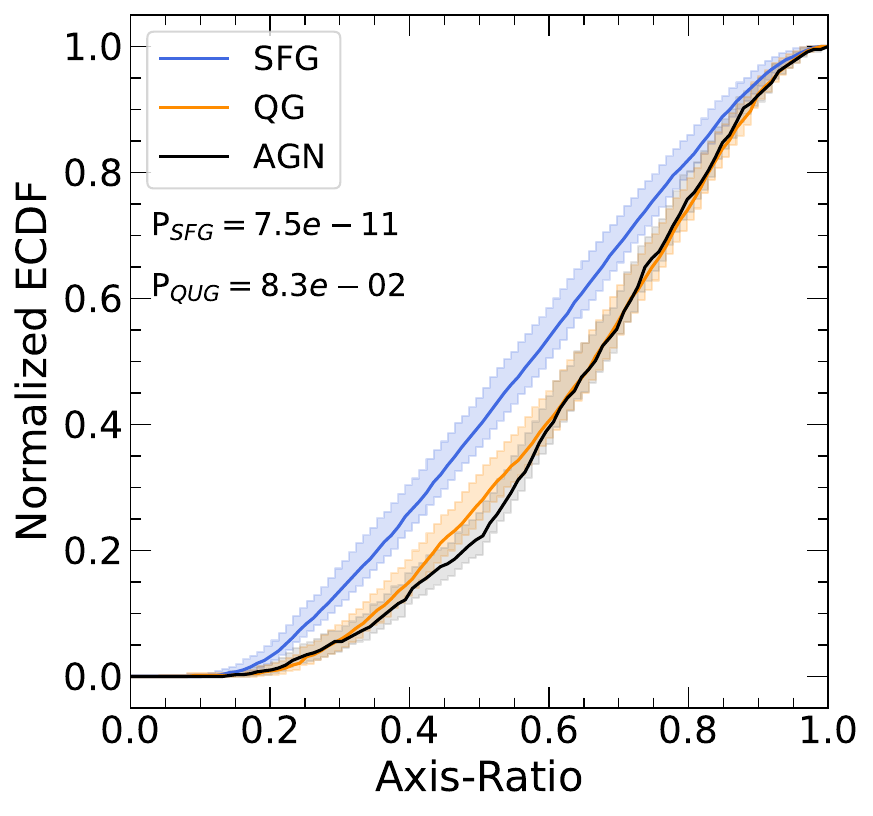}
    \caption{Normalized cumulative distribution of the projected axis ratio. AGN host, SFG, and QG are shown in black, blue, and orange lines, respectively. The shared area shows the 16th and 84th percentiles of the distribution. The axis ratio distribution of AGN is similar to QGs rather and SFG}
    \label{fig:axdist}
\end{figure}

Here, we do not bin the sample in redshift or stellar mass, but we perform the redshift-stellar mass distribution matching directly at once. We construct 100 mock AGN samples with 100 mock SFG and QG samples each. The axis ratio distribution of the AGN sample against SFG and QG is shown in figure~\ref{fig:axdist}. The axis ratio of quiescent galaxies is larger than SFG. This may be due to an intrinsic difference in the shape of SFG and QG, where SFG are disk-like galaxies while QG are spheroidal or elliptical.  For AGN host galaxies, the axis ratio is comparable to that of quiescent galaxies and shows a strong difference from star-forming galaxies. The test  indicates that the axis-ratio distribution of AGN host galaxies differs strongly from SFG. However, against QG, we cannot reject the null-hypothesis that the axis-ratio distribution of AGN host differs from QG are different at a 99\% confidence (P-value 0.083). We conclude that the axis-ratio of AGN host galaxies is similar to QGs.

\subsection{1-\um Size-Mass Relationship of AGN host} \label{sec:res:sizemass}

\begin{figure*}[ht]
    \centering
    \includegraphics[width=\linewidth]{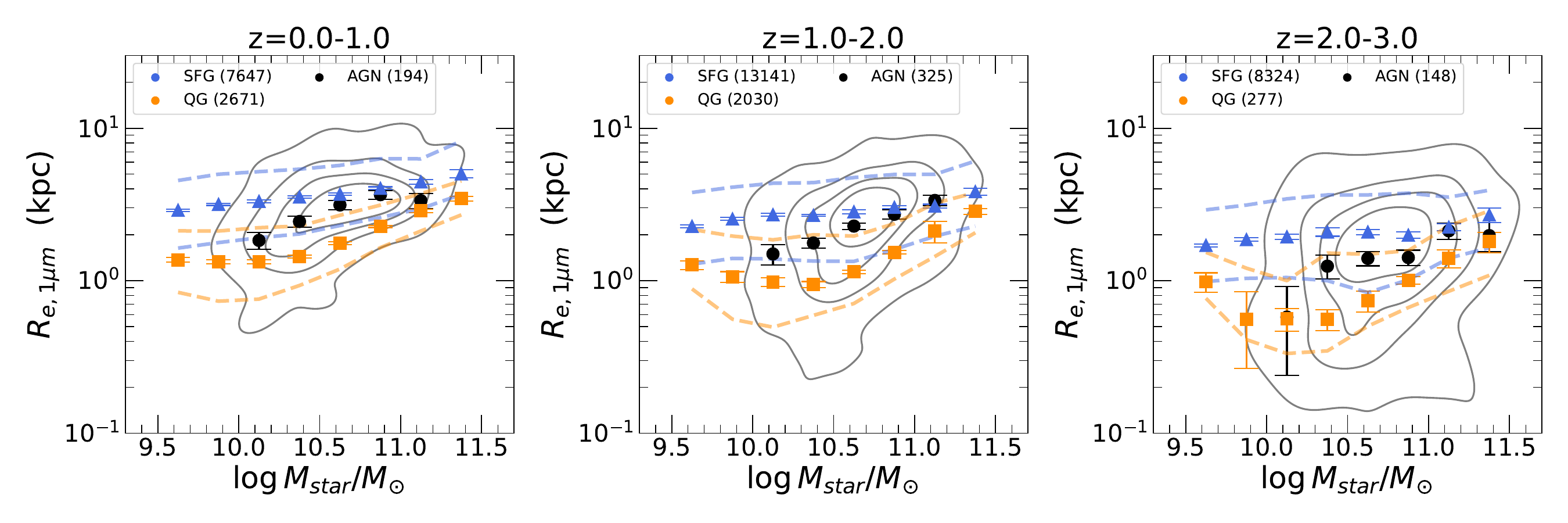}
    \caption{Binned size-mass relation of AGN host, SFG, and QG shown in black, blue, and orange dots, respectively. The dotted line shows the 16th and 84th percentile distribution of the samples. The uncertainties represent the standard error of the mean. The gray contours show the distribution of the AGN host galaxies.}
    \label{fig:re_bin_size}
\end{figure*}

After estimating the 1-\um rest-frame effective radius of the AGN host galaxy, we compared the size of the AGN host galaxy to the SFG and QG samples. First, we estimate the median 1-\um size by binning the samples in redshift and stellar mass. Figure~\ref{fig:re_bin_size} shows the binned effective radius of AGN host galaxies compared with SFG and QG. In all redshift bins, AGN host galaxies are smaller than star-forming galaxies but remain larger than quiescent galaxies. In the QG sample, we observe a flattening of the sizes at stellar mass below  $\log M_s \sim 10.3$. 

\begin{figure*}[htb]
    \centering
    \includegraphics[width=\linewidth]{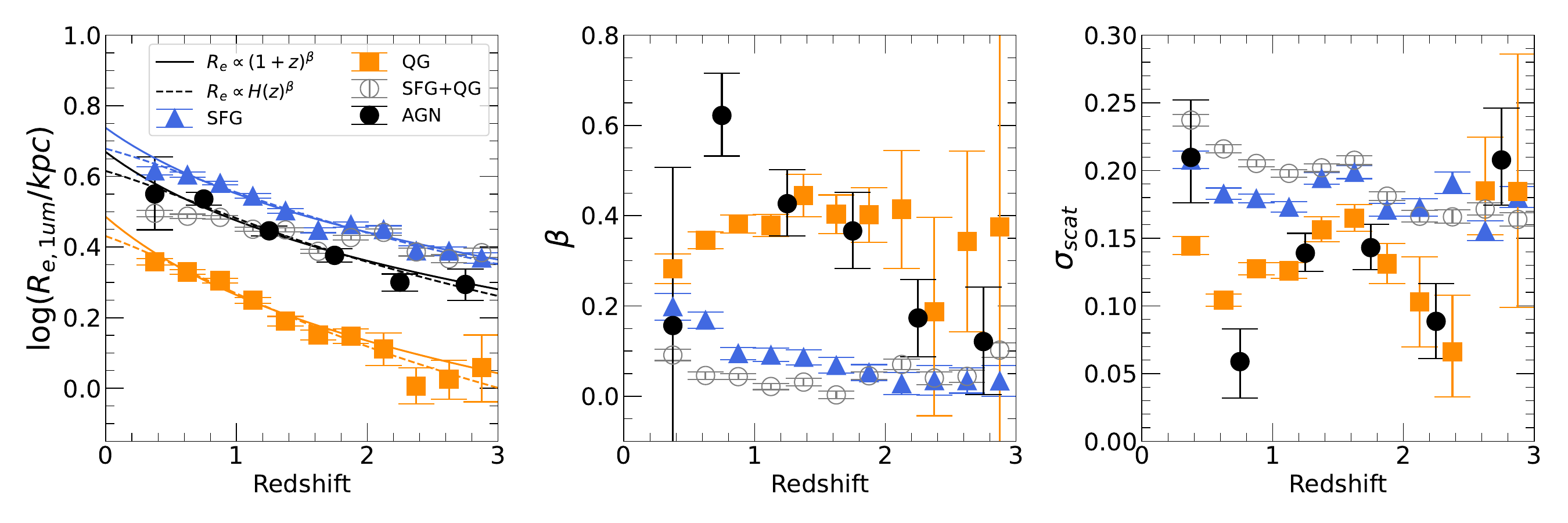}
    \caption{The average size ($\log R_{e,1\mu m}$), slope ($\beta$) , and $1\sigma$ scatter of the best-fitted size-mass relationship. The parameters for AGN host, SFG, QG, and SFG+QG are shown in black round, blue triangles, orange boxes, and open round symbols, respectively. The sizes of AGN host galaxies follow the same trend as QGs and SFG, but with a size between the two populations. The slope of the AGN size mass relationship is, on average, larger than SFGs.}
    \label{fig:sm_fit_par}
\end{figure*}

To investigate the evolution of the size-mass relationship as a function of redshift, we parameterize the size-mass relationship as a linear function in each redshift bin following \citet{2014ApJ...788...28V} as 

\begin{equation}
    \log R_{\rm e}(M_s) = \log R_{\rm e, 0} + \beta \log\bigg( \frac{M_{s}}{5\times 10^{10} M_{\odot}} \bigg)
\end{equation}

where $\log R_{\rm e}$ is the restframe effective radius, $\log R_{\rm e,0}$  is the effective radius at stellar mass of $M_s=5\times 10^{10} M_\odot$, and $\beta$ is the slope of the size mass relationship. We used the Bayesian linear regression code \citep{2007ApJ...665.1489K}, which also provides the $1\sigma$ scatter ($\sigma_{\rm scat}$) of the data to perform the fitting and takes into account the uncertainty of both the stellar mass and size. We also include an additional systematic uncertainty based on the FWHM of the pivot band ($\sigma_{R_{e,sys}}=\rm FWHM_{pivot}/2.355$). It depends on the band used to estimate the size of the galaxy and its redshift.  We limit our analysis to AGN and SFG with a stellar mass between $\log M_{s}/M_\odot=10-11.5$, which is the stellar mass where most of the AGN sample falls. For QG, we limit the stellar mass to between $\log M_{s}/M_\odot=10.3-11.5$, slightly higher than SFG. This limit is set to prevent artificially overestimating the size of QG due to the flattening of the QG mass-size slope\citep{2014ApJ...788...28V,2021ApJ...921...38K,2024ApJ...960...53V,2024A&A...691A.240M}, which appears to flatten out below $\log M_s<10.3$ in our data. We first perform the fitting to the AGN host sample, limiting the sample to 6 redshift bins between redshift 0 to 3 due to the smaller sample size. For the comparison sample, we perform the fitting in smaller redshift bins as the number of SFG and QG is much larger than the AGN host galaxies. The best-fitted parameters are shown in Table~\ref {tab:smpar_all}. The best-fit parameters of the size mass relationship are shown in Figure~\ref {fig:sm_fit_par}.

Our results  show that AGN host galaxies are smaller towards the early universe, a trend seen similarly with star-forming and quiescent galaxies. The effective radius is smaller than SFGs but still larger than QGs. Previous studies also investigate the evolution of the size-mass relationship by parameterizing the size evolution as a function of $R_{\rm e,0} = \alpha(1+z)^\beta$ and  $R_{\rm e,0} = \alpha H(z)^\beta$ where $\beta$ is the slope of the size evolution and $H(z)$ is the Hubble constant at that redshift. For comparison, we also perform the fitting with the best-fit parameters are shown in table~\ref{tab:size_evopar}, which shows that for all 3 populations, the size evolution is better described with  $R_{\rm e,0}  \propto H(z)^\beta$ with $\beta\sim -0.6$. As for the slope, on average slope of the size mass relationship of AGN is steeper than the slope of the  SFG size mass relationship. Our measurements suggest a flattening of the slope towards the early universe. However, we conclude that the evolution in the slope of the size-mass relationship of AGN host galaxies remains inconclusive due to limitations in the sample size and mass coverage. 
    
\begin{deluxetable}{cccc}[!htb]
    \tablecaption{Normalization, slope of the size evolution and $\chi_{\nu}^2$}
    \label{tab:size_evopar}
    \tablehead{ Sample & {$\log \alpha$ } & $\beta$ & $\chi_{\nu}^2$ }
    \startdata
    \multicolumn{4}{c}{$R_{\rm e,0} = \alpha (1+z)^\beta$} \\ 
    \hline
    SFG & $0.74\pm0.02$ & $-0.62\pm0.05$ & 4.78 \\
    QG & $0.49\pm0.02$ & $-0.73\pm0.07$	& 3.03 \\ 
    AGN & $0.67\pm0.04$ & $-0.65\pm0.09$ & 1.20 \\ 
    \hline 
    \multicolumn{4}{c}{$R_{\rm e,0} = \alpha H(z)^\beta$} \\ 
    \hline 
    SFG & $1.61\pm0.08$	& $-0.50\pm0.04$ & 4.06 \\ 
    QG  & $1.65\pm0.09$ & $-0.66\pm0.04$ & 1.66 \\ 
    AGN & $1.62\pm0.08$ & $-0.55\pm0.04$ & 0.24 \\
    \hline
\enddata
\end{deluxetable}
 
\begin{figure}[htb!]
    \centering
    \includegraphics[width=\linewidth]{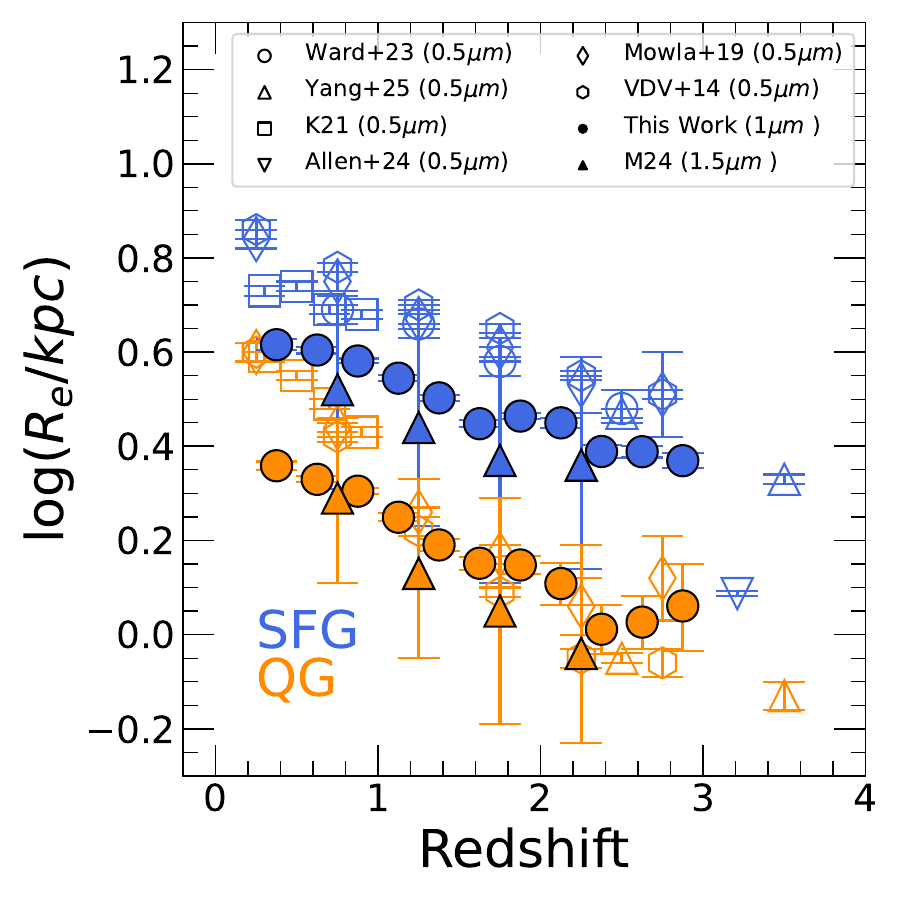}
    \caption{Effective radius evolution of SFG and QG. Our results at 1\um are shown with filled round symbols.  Near-infrared sizes at 1.5 \um from \citealt{2024ApJ...972..134M} are shown with filled triangles. Optical 0.5 \um sizes from \citealt{2014ApJ...788...28V,2019ApJ...880...57M,2021ApJ...921...38K,2024arXiv241016354A,2024ApJ...962..176W} and \citealt{2025arXiv250407185Y} are show with open symbols. }
    \label{fig:05sizemass}
\end{figure}

Figure \ref{fig:05sizemass} shows the average size of our comparison galaxies at 1\um with previous measurements at the 0.5\um wavelength \citealt{2014ApJ...788...28V,2019ApJ...880...57M,2021ApJ...921...38K,2024arXiv241016354A,2024ApJ...962..176W,2025arXiv250407185Y} and the 
near-infrared size (1.5 \um, \citealt{2024ApJ...972..134M}) from the literature.  We find the same trends as those described in the literature. For the same stellar mass, SFGs are larger than QGs, and on average, both of them are smaller at high redshift. The slope of the size mass relationship of QG is steeper than SFGs. There is a strong wavelength dependence from the optical to near infrared, where the redder the bands, the smaller the effective radius is. This effect appears stronger in SFG than QG due to the larger size gradient on the massive end, as shown in section~\ref{sec:anal:resteffrad}.

\begin{deluxetable}{ccccc}[!htb]
    \tablecaption{Size-Mass Relationship Coefficients of SFG and QG}
    \label{tab:smpar_all}
    \tablehead{$z$ & N & $\log (R_{\rm e,0}/\rm kpc)$ & $\beta$ & $\epsilon$}
    \startdata
    \multicolumn{5}{c}{  AGN } \\
    0.25-0.5 & 24 & $0.55_{-0.10}^{+0.10}$ & $0.15_{-0.35}^{+0.33}$ & $0.21_{-0.03}^{+0.04}$ \\
    0.5-1.0 & 169 & $0.54_{-0.02}^{+0.02}$ & $0.62_{-0.08}^{+0.10}$ & $0.06_{-0.02}^{+0.02}$ \\
    1.0-1.5 & 186 & $0.45_{-0.01}^{+0.01}$ & $0.43_{-0.07}^{+0.07}$ & $0.14_{-0.01}^{+0.01}$ \\
    1.5-2.0 & 139 & $0.38_{-0.02}^{+0.02}$ & $0.37_{-0.08}^{+0.08}$ & $0.14_{-0.02}^{+0.02}$ \\
    2.0-2.5 & 85 & $0.30_{-0.03}^{+0.02}$ & $0.17_{-0.09}^{+0.09}$ & $0.09_{-0.03}^{+0.03}$ \\
    2.5-3.0 & 63 & $0.29_{-0.05}^{+0.04}$ & $0.12_{-0.12}^{+0.12}$ & $0.21_{-0.03}^{+0.04}$ \\
    \hline
    \multicolumn{5}{c}{ SFG } \\
    \hline
    0.25-0.5 & 563 & $0.62_{-0.01}^{+0.01}$ & $0.20_{-0.03}^{+0.03}$ & $0.21_{-0.01}^{+0.01}$ \\
    0.5-0.75 & 1205 & $0.60_{-0.01}^{+0.01}$ & $0.17_{-0.02}^{+0.02}$ & $0.18_{-0.01}^{+0.01}$ \\
    0.75-1.0 & 1959 & $0.58_{-0.01}^{+0.01}$ & $0.09_{-0.01}^{+0.01}$ & $0.18_{-0.01}^{+0.01}$ \\
    1.0-1.25 & 1493 & $0.54_{-0.01}^{+0.01}$ & $0.09_{-0.01}^{+0.02}$ & $0.17_{-0.01}^{+0.01}$ \\
    1.25-1.5 & 1622 & $0.50_{-0.01}^{+0.01}$ & $0.09_{-0.02}^{+0.02}$ & $0.19_{-0.01}^{+0.01}$ \\
    1.5-1.75 & 1397 & $0.45_{-0.01}^{+0.01}$ & $0.07_{-0.02}^{+0.02}$ & $0.20_{-0.01}^{+0.01}$ \\
    1.75-2.0 & 1651 & $0.46_{-0.01}^{+0.01}$ & $0.05_{-0.02}^{+0.02}$ & $0.17_{-0.01}^{+0.01}$ \\
    2.0-2.25 & 996 & $0.45_{-0.01}^{0+.01}$ & $0.03_{-0.02}^{+0.02}$ & $0.17_{-0.01}^{+0.01}$ \\
    2.25-2.5 & 781 & $0.39_{-0.01}^{+0.01}$ & $0.03_{-0.03}^{+0.03}$ & $0.19_{-0.01}^{+0.01}$ \\
    2.5-2.75 & 817 & $0.39_{-0.01}^{+0.01}$ & $0.03_{-0.03}^{+0.03}$ & $0.16_{-0.01}^{+0.01}$ \\
    2.75-3.0 & 656 & $0.37_{-0.02}^{+0.02}$ & $0.03_{-0.03}^{+0.03}$ & $0.18_{-0.01}^{+0.01}$ \\
    \hline 
    \multicolumn{5}{c}{ QG } \\
    \hline
    0.25-0.5 & 279 & $0.36_{-0.01}^{+0.01}$ & $0.28_{-0.03}^{+0.03}$ & $0.14_{-0.01}^{+0.01}$ \\
    0.5-0.75 & 568 & $0.33_{-0.01}^{+0.01}$ & $0.35_{-0.02}^{+0.02}$ & $0.10_{-0.01}^{+0.01}$ \\
    0.75-1.0 & 822 & $0.30_{-0.01}^{+0.01}$ & $0.38_{-0.02}^{+0.02}$ & $0.13_{-0.01}^{+0.01}$ \\
    1.0-1.25 & 593 & $0.25_{-0.01}^{+0.01}$ & $0.38_{-0.02}^{+0.02}$ & $0.13_{-0.01}^{+0.01}$ \\
    1.25-1.5 & 317 & $0.19_{-0.01}^{+0.01}$ & $0.44_{-0.05}^{+0.05}$ & $0.16_{-0.01}^{+0.01}$ \\
    1.5-1.75 & 372 & $0.15_{-0.01}^{+0.01}$ & $0.40_{-0.04}^{+0.04}$ & $0.16_{-0.01}^{+0.01}$ \\
    1.75-2.0 & 270 & $0.15_{-0.02}^{+0.02}$ & $0.40_{-0.06}^{+0.06}$ & $0.13_{-0.01}^{+0.01}$ \\
    2.0-2.25 & 77 & $0.11_{-0.05}^{+0.04}$ & $0.43_{-0.13}^{+0.13}$ & $0.10_{-0.03}^{+0.03}$ \\
    2.25-2.5 & 55 & $0.01_{-0.05}^{+0.05}$ & $0.16_{-0.21}^{+0.23}$ & $0.06_{-0.03}^{+0.04}$ \\
    2.5-2.75 & 67 & $0.03_{-0.06}^{+0.05}$ & $0.34_{-0.19}^{+0.20}$ & $0.19_{-0.03}^{+0.04}$ \\
    2.75-3.0 & 26 & $0.06_{-0.10}^{+0.09}$ & $0.33_{-0.66}^{+0.65}$ & $0.18_{-0.09}^{+0.10}$ \\
    \hline
\enddata
\end{deluxetable}

\subsection{Non-parametric morphologies} \label{sec:res:npar}

One of the greatest strengths of non-parametric morphologies is their ability to classify galaxy mergers and shapes. Here, we estimate the fraction of mergers among our AGN host galaxy sample and compare it with the comparison galaxies. Since morphologies can be dependent on stellar mass, we again perform the stellar mass and redshift matching described in section \ref{sec:res:Sersic} before estimating the fraction of mergers, spheroidal galaxies, and spiral galaxies.

\begin{figure*}
    \centering
    \includegraphics[width=\linewidth]{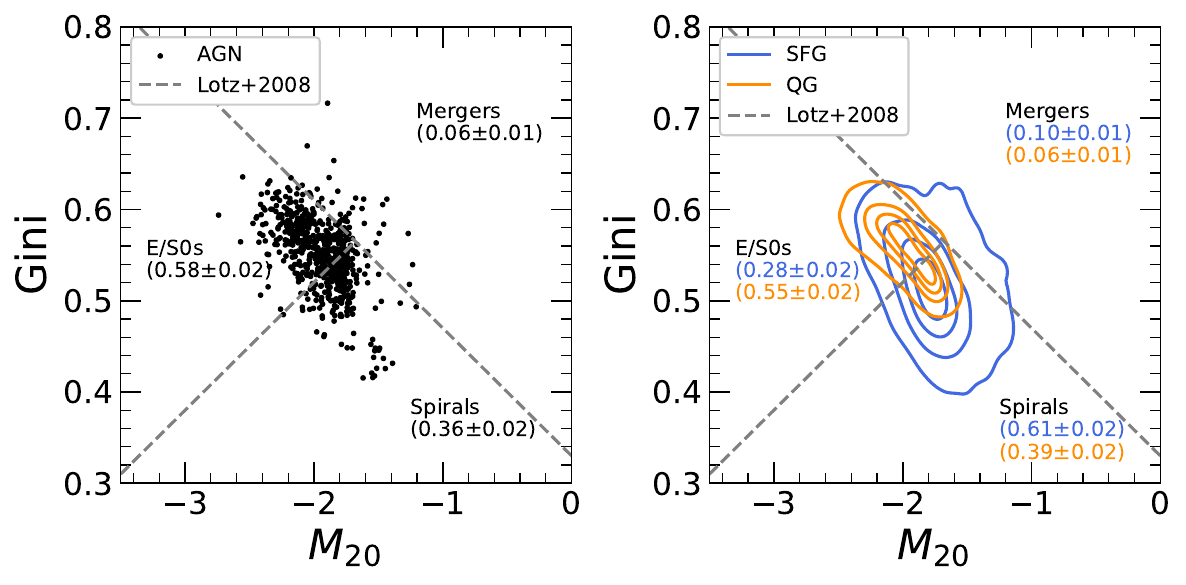}
    \caption{Gini-$M_{20}$ distribution of AGN host galaxies (Left)  and the SFG and QG (Right). The dashed lines show the boundaries for Merger, spheroidal, and spiral galaxy classifications described in \citet{2008ApJ...672..177L}. Most of the X-ray selected AGN are not classified as mergers. }
    \label{fig:nonpar}
\end{figure*}

Figure \ref{fig:nonpar} shows the distribution of AGN host galaxies, SFG, and QG on the Gini-$M_{20}$ diagram. Most of the AGN host galaxies are classified as non-merger galaxies. This indicates that most of the AGN hosts have a non-disturbed morphology. Only 6\%  of the AGN host galaxies are classified as mergers as opposed to the 6-10\% seen in QG and SFGs. Among the AGN host galaxies, 57\% of them have spheroidal morphologies and 37\% are spiral galaxies. This fraction is close to that seen in the QG comparison sample, which is 55\% spheroidal and 39\% spiral. SFG has approximately a factor of 2 larger galaxies classified as spirals.  This result is overall consistent with the Sersic index distribution, which has a mean of $n_s\sim 2$ for the AGN host galaxies, which  implies a spheroidal component within the host. Interestingly, the fraction of mergers in the AGN host sample is lower than SFGs. This could indicate that the AGN host galaxies are already post-mergers and their morphological transformation has progressed enough towards spheroidal morphologies. We will discuss this further in section \ref{sec:discussion}. 

\section{Discussion} \label{sec:discussion}

\subsection{AGN Activity \& Morphology} \label{sec:dis:evolcomp}

 In section~\ref{sec:res:Sersic}, we showed that AGN host galaxies have Sersic indexes between QGs and SFGs across redshift and stellar mass bins. AGN host galaxies also have axis ratios similar to QGs but differ from SFGs. From the point of view of non-parametric morphologies, most AGN have shapes akin to spheroidal galaxies with a similar fraction to QGs as opposed to SFGs, which are mostly spiral-like galaxies. 
 
 Through the size-mass relationship, we also showed that the size of AGN host galaxies is between SFGs and QGs, but the slope of their size-mass relation is steeper than SFGs. The AGN host galaxies are also smaller at higher redshift, following similar trends to SFG and QGs. They also have a size gradient where the size of shorter wavelengths is larger than redder wavelengths. If we take at face value the NUVrJ classification of AGN host which suggest that our sample of AGN host composed of mostly SFGs, then our results are consistent with findings of the size of AGN host galaxies \citep{2019ApJ...887L...5S,2021ApJ...918...22L,2024MNRAS.527.4690L,2024ApJ...962...93Z}, which also find AGN host galaxies within the intermediate size range between star-forming galaxies and quiescent galaxies. 
 
 Overall, our results indicate that the AGN host galaxies have a significant spheroidal component or are undergoing morphological transformation by building up the bulge, since the residuals of the AGN host galaxy fitting also show spiral arms and not a pure bulge structure. This fits with the model that AGN activity can occur simultaneously with morphological transformation and is a transitional phase between the two.

\subsection{Secular or Merger-driven AGN Triggering} \label{sec:dis:trig}

Previous work that investigates AGN triggering by investigating the host galaxies tends to find different results due to sample selection. On the one hand, studies of high-redshift AGN host galaxies, including quasars and X-ray selected AGN, prefer secular triggers of AGN activity, such as violent disk instabilities \citep{2006ApJS..166....1H,2011ApJ...741L..33B}. On the other hand, infrared-selected AGN prefer to be merger-triggered \citep{2006ApJS..163....1H,2018MNRAS.478.3056B}. The most prominent case of merger-triggered AGN is local Ultra Luminous Infrared Galaxies (ULIRGS \cite{1996ARA&A..34..749S}) which most are powered by AGN and tend to possess tidal features and amorphous morphology indicative of recent wet major mergers \citep{2008PASJ...60S.489I,2011MNRAS.410..573G,2017ApJ...835...36T,2014ApJ...794..139I,2017MNRAS.468.1273R,2021MNRAS.506.5935R,2021ApJS..257...61Y,2022ApJ...936..118Y,2022PASJ...74.1356T}. Recently, \citet{2023OJAp....6E..34V} perform an analysis of the merger fractions across AGN selection and luminosity and conclude that the difference is not due to observational bias between datasets or systematics.

\begin{figure*}[htb]
    \centering
    \includegraphics[width=\linewidth]{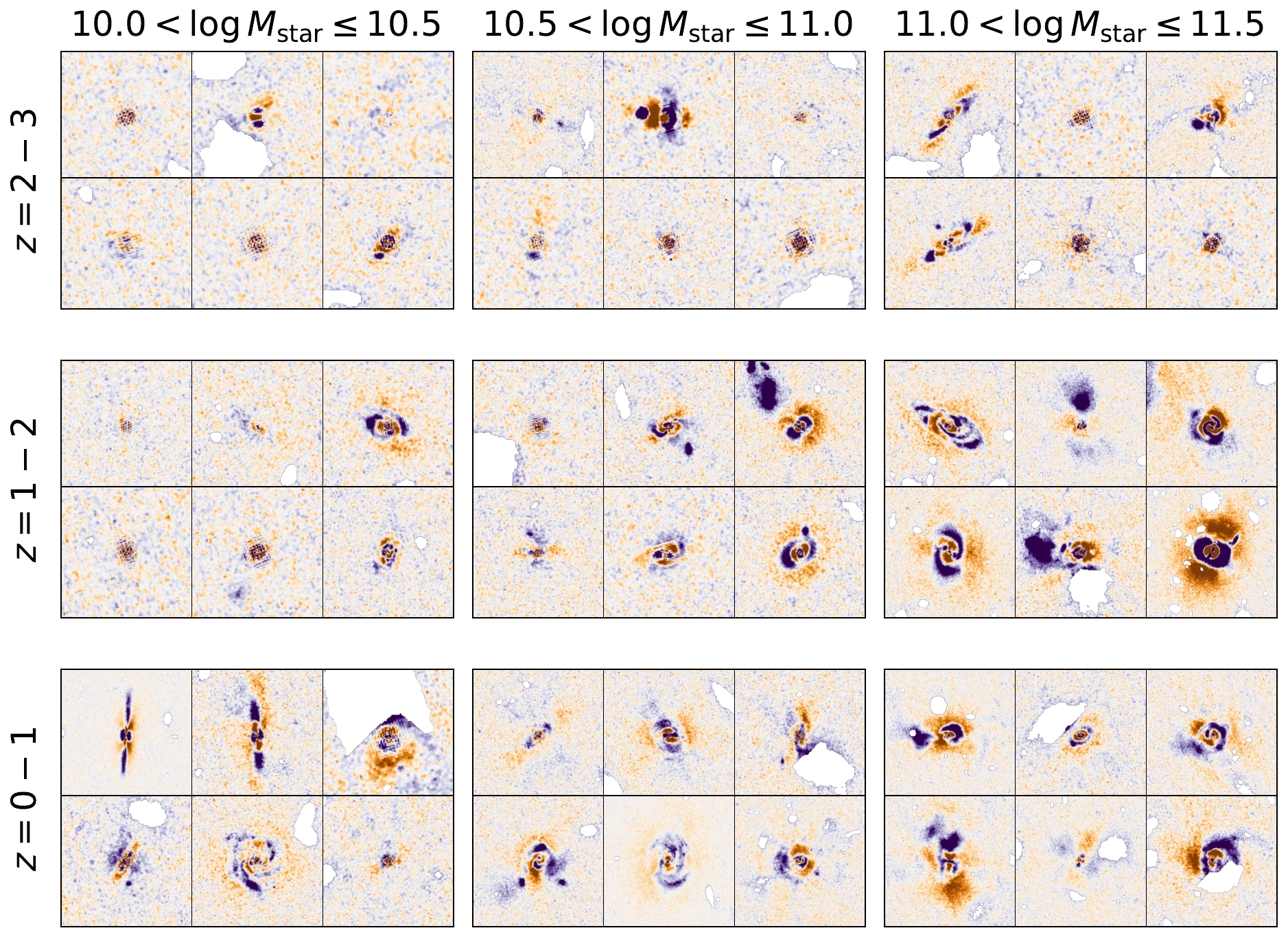}
    \caption{Randomly selected uncertainty weighted residual images of the AGN host galaxies. The residuals are grouped by stellar mass and redshift. The images correspond to the pivot image used to calculate the size of the AGN host galaxies. Orange and Purple colors represent negative and positive residuals, respectively.}
    \label{fig:substruc}
\end{figure*}

Our analysis of non-parametric morphologies shows a low fraction ($6\%$) of merger morphologies among AGN host galaxies and compared to SFG ($\sim 10\%$).  The Sersic indices, axis ratios, and non-parametric classification of AGN host galaxies indicate that many AGN in our sample could be post-major merger galaxies with spheroidal morphologies. However, features consistent with spiral arms and bars are seen in the residual images of figure~\ref{fig:substruc}. These substructures are also seen in previous work using JWST data \citep{2025ApJ...979..215T,2024ApJ...962...93Z} and are strongly in tension with evolutionary pathways that invoke major mergers as a ubiquitous trigger for the morphological transformation and AGN activity. Due to the strong gravitational torques that destroy spiral or bar structures, disturbed or spheroidal morphology is expected during late and post-merger (eg. \citealt{2008ApJS..175..356H,2008ApJS..175..390H}). Hence, at face value, our results would indicate that most AGN in our sample are secular triggering. 

However, the situation appears to be more complicated when considering the time scales. Recently, \citet{2025ApJ...989...73O}  found none to weak enhancements in AGN and star-formation activity among a sample of merging galaxies. They discuss that the weak enhancement is due to the different levels of star-formation and AGN activity observed in each merger stage. The different timescales of merger features, star-formations, and AGN activity combined weaken the statistical evidence of AGN \& star-formation enhancement in mergers. \citet{2025ApJ...978...74B} used the asymmetry parameters to show that many obscured AGN are disturbed, especially heavily obscured AGN, This is in tension with previous observations (eg. \citealt{2011ApJ...727L..31S,2011ApJ...726...57C,2012ApJ...744..148K,2019ApJ...877...52Z}),  which show that most moderately luminous AGN (such as in this work) are not disturbed galaxies.  This may also be a result of the timescales at which merger signatures can be observed. In the case of the Gini-$M_{20}$, the observability being a merger is $\sim 0.3$ Gyr \citep{2008MNRAS.391.1137L,2010MNRAS.404..590L,2010MNRAS.404..575L} whereas asymmetry has a longer time scale $\sim 1$ Gyr \citep{2010MNRAS.404..575L,2017MNRAS.465.1106B}.  High-z compaction event can also change the Sersic index of galaxy from 1.5 to $\sim 4$ within 0.6 Gyr \citep{2016MNRAS.458..242T}. 

\begin{figure}[htb]
    \centering
    \includegraphics[width=\linewidth]{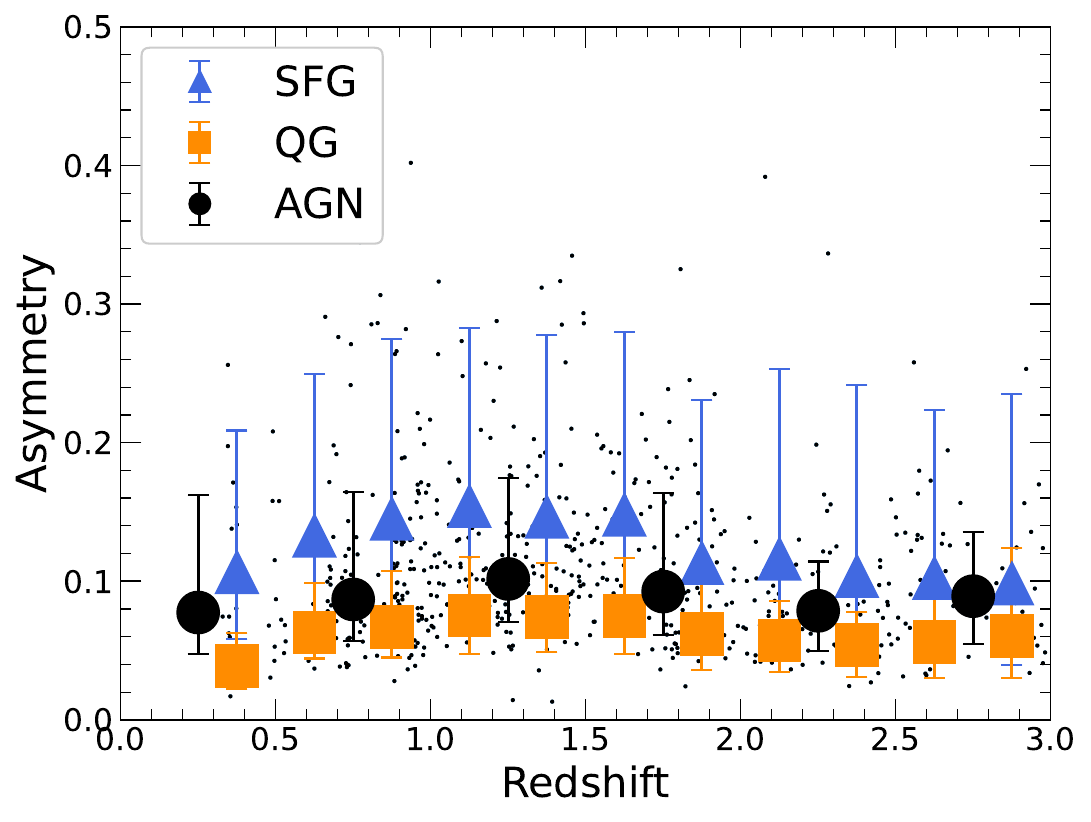}
    \caption{Median asymmetry of AGN (round black symbols), SFG (blue triangles), and QG (orange squares) in redshift bins between redshift 0-3.  The lower and upper uncertainties are the 16th and 84th percentiles, respectively.  The small black dots are AGN host galaxies. The asymmetry has a flat evolution with redshift and a value of $\sim 0.1$}
    \label{fig:asym}
\end{figure}

The contradiction between studies may also be driven by the matching between the observability window of merger classification and the windows of AGN selection.  Typically, the average AGN duty cycle is a few 100 Myrs \citep{1982MNRAS.200..115S,2002MNRAS.335..965Y,2004MNRAS.351..169M}. However, recent simulations of galaxy mergers by \citet{2022ApJ...936..118Y} showed that the windows during which merging galaxies can be selected as AGN-dominated dust-obscured galaxies (also known as HotDogs) using infrared color selection are approximately $\sim$ 4 Myrs, which is much shorter. \citet{2016ApJ...822L..32F} showed that HotDOGs  have a merger fraction as high as $62\pm14\%$. This means that infrared selection is not only sensitive to obscured sources but also more sensitive to mergers than other selection methods, such as X-ray selection. If so, our sample may contain mainly pre-merger absorbed and unabsorbed AGN (disk or spiral host) and post-merger unabsorbed AGN (spheroidal host), as we miss heavily obscured merger stages. 
These observations are also in line with the results of \citet{2024MNRAS.533.3068B}, who recently also concluded that X-ray selected AGN are more likely to be selected in a pre-merger phase. The asymmetry parameter of AGN host galaxies compared to SFG and QG is shown in  Figure~\ref{fig:asym}. Our asymmetry parameters using the same bands and the non-parameter (close to 1 \um) have a flat evolution with redshift, with a median value of $\sim 0.1$, which is lower than what \citet{2025ApJ...978...74B} found for heavily obscured sources. However, the lower asymmetry parameter may be due to masking of nearby sources during the processing of our sample.

Besides timescales, the gas-rich environment may also affect morphologies post major mergers. Recent simulations of high redshift galaxy mergers showed that gas-rich mergers can regrow their disk and spiral arms in gas-rich environments such as the early universe \citep{2009ApJ...691.1168H,2018MNRAS.480.2266M}. This could explain the observed spiral arms and bars we observe in the residual image.  These simulations challenge the paradigm that moderate luminosity AGN may be preferentially triggered by violent disk instabilities, whereas high luminosity AGN may be triggered by mergers \citep{2012ApJ...751...72D,2014A&A...569A..37M}. It is currently unclear which pathway dominates the morphological transformation and triggering of AGN host galaxies. In reality, the morphological transformation of galaxies and the triggering of AGN can likely occur through both secular and merger pathways. 

\section{Summary} \label{sec:summary}

In this work, we leveraged the high angular resolution of JWST to investigate the host galaxy morphology of moderate luminosity X-ray selected AGN within the COSMOS field. We fitted the sample of AGN host galaxies between redshift 0-3 with 2D Sersic models and performed host galaxy decomposition to account for unobscured AGN light if required. The comparison galaxies were fitted with a single Sersic component. Using the AGN-subtracted photometry, we perform SED fitting using the available JWST bands to estimate the stellar mass of the AGN host galaxy. We constructed the size-mass relationship for AGN host galaxies between redshift 0-3 and performed statistical analysis of the parametric and non-parametric shapes of the AGN host galaxy compared to non-active galaxies.

\begin{enumerate}
    \item The AGN have Sersic indices $\sim 2$ which is in between star-forming galaxies ($\sim 1.5$) and quiescent galaxies ($>3$). The Sersic index of all three samples increases with stellar mass.
    \item The distribution of AGN host galaxies' axis-ratio is significantly different from star-forming galaxies but similar to quiescent galaxies. The distribution of the sersic index among AGN host galaxies differs from SFGs and QGs. 
    \item The average size of AGN host galaxies below redshift 3 is 1-5 kpc, in between the SFG and QG populations. They follow the same trend as SFG and QG, increasing size with stellar mass, but are smaller towards high redshift for the same stellar mass. 
    \item On average, the AGN size-mass relationship has a slope of ($\beta_{\rm AGN}\sim 0.35$), but the evolution of the slope is still unclear due to limited coverage in stellar mass. This slope is larger than  ($\beta_{SFG} \sim 0.1$) and closer to QGs ($\beta_{QG}\sim 0.4$).  
    \item The size evolution of AGN, SFG, and QG host galaxies is better described as a function of $H(z)$ with a slope of $\beta \sim -0.6$.     
    \item Non-parametric morphologies indicate that most AGN host galaxies have shapes similar to spheroidal galaxies ($\sim 57\%$) and only 37\% are spiral galaxies. The fraction of merger hosts is small, only 6\% but can be affected by selection effects.
\end{enumerate}

Overall, our results show that AGN activity can occur simultaneously with morphological transformation, as the triggering mechanism is similar. The shapes and size suggest that the transformation towards spheroidal morphologies has significantly progressed, given the Sersic index, which indicates significant spheroidal components.  Additional high angular resolution observations of AGN host galaxies over a broad AGN luminosity, redshift, and stellar mass range would help differentiate pathways of the morphological transformation of galaxies.  We will provide all size and shape measurements of AGN host galaxies, SFGs, and QGs, as well as future revisions at  \dataset[doi:10.5281/zenodo.17176457]{https://doi.org/10.5281/zenodo.17176457}.

\section*{acknowledgments}
We thank the anonymous reviewer for their helpful comments which improved the draft. We acknowledge support from the National Science and Technology Council of Taiwan grants 111-2112-M-001-052-MY3 (WHW, BV, ZKG) and 112-2811-M-001-129-MY2 (BV). C.-C.C. and acknowledges support from the National Science and Technology Council of Taiwan (111-2112-M-001-045-MY3), as well as Academia Sinica through the Career Development Award (AS-CDA-112-M02). The SuMIRe Cluster is primarily funded by the Academia Sinica Investigator Award (grants AS-IA-107-M01 and AS-IA-112-M04) and the National Science and Technology Council, Taiwan (grant NSTC 112-2112-M-001-027-MY3). The authors acknowledge the access to high-performance computing facilities (Theory cluster and storage) provided by Academia Sinica Institute of Astronomy and Astrophysics (ASIAA). This research has made use of the SAO/NASA Astrophysics Data System. All of the data presented in this article were obtained from the Mikulski Archive for Space Telescopes (MAST) at the Space Telescope Science Institute. The specific observations analyzed can be accessed via \dataset[doi:10.17909/a0kd-c157]{https://doi.org/10.17909/a0kd-c157} (COSMOS-Web) and \dataset[doi:10.17909/zfgy-d978]{https://doi.org/10.17909/zfgy-d978} (PRIMER)."

\bibliography{sample631}{}
\bibliographystyle{aasjournal}



\end{document}